\pdfoutput=1
\documentclass[11pt]{article}

\PassOptionsToPackage{table}{xcolor}
\usepackage[final]{acl}

\usepackage{url}
\usepackage{times}
\usepackage{latexsym}
\usepackage[T1]{fontenc}
\usepackage[utf8]{inputenc}
\usepackage{microtype}
\usepackage{inconsolata}
\usepackage{graphicx}
\usepackage{booktabs}
\usepackage{multirow}
\usepackage{array}
\usepackage{xcolor}
\usepackage{amsmath}
\usepackage{enumitem}
\usepackage{makecell}
\usepackage{tikz}
\usetikzlibrary{positioning,arrows.meta,calc,fit,backgrounds}
\definecolor{paperbg}{HTML}{F5F1E8}
\definecolor{cardbg}{HTML}{FBF8F1}
\definecolor{iobg}{HTML}{EDE6D3}
\definecolor{stroke}{HTML}{C9C0AC}
\definecolor{ink}{HTML}{1F1F1F}
\definecolor{muted}{HTML}{6B6256}
\definecolor{accentblue}{HTML}{2E5C8A}
\definecolor{accentorange}{HTML}{B8731E}
\definecolor{accentpurple}{HTML}{7A3A5C}
\definecolor{accentgreen}{HTML}{4A7A4A}
\definecolor{okbg}{HTML}{D5E5D0}

\definecolor{domainbest}{HTML}{CFE2F3}  
\definecolor{overallbest}{HTML}{FCE4D6} 

\definecolor{oursrow}{HTML}{E2F0D9}      
\definecolor{baselinerow}{HTML}{FCE4D6}  

\newcommand{\ourscell}[1]{\cellcolor{oursrow}\textbf{#1}}
\newcommand{\basecell}[1]{\cellcolor{baselinerow}#1}
\newcommand{\bestdom}[1]{\cellcolor{domainbest}\textbf{#1}}
\newcommand{\bestavg}[1]{\cellcolor{overallbest}\textbf{#1}}
\newcommand{\midr}{\textsc{MIDR}}

\newcommand{\ndcg}{nDCG@10}

\usepackage{float}                  
\usepackage[most]{tcolorbox}        
 
\newtcolorbox{fullpromptcard}[1]{%
  breakable,
  enhanced,
  colback=cardbg,
  colframe=stroke,
  coltitle=ink,
  colbacktitle=iobg,
  boxrule=0.4pt,
  arc=2pt,
  left=6pt, right=6pt, top=4pt, bottom=4pt,
  fonttitle=\bfseries\small,
  title={#1},
  fontupper=\ttfamily\scriptsize,
  before upper={\parindent0pt\setlength{\parskip}{2pt}},
}

\newenvironment{promptfigure}
  {\par\addvspace{\intextsep}}
  {\par\addvspace{\intextsep}}

\title{MIDR: Enrichment-Augmented Indexing for Multimodal Document Retrieval}

\author{
  \textbf{Debanjan Mahata},
  \textbf{Atharva Tendle},
  \textbf{Daniel Preo\c{t}iuc-Pietro},
  \textbf{Yong Zhuang},
  \textbf{Ozan \.{I}rsoy}
\\
  Bloomberg, NYC, USA
\\
  \small{
    \texttt{\{dmahata,atendle,dpreotiucpie,yzhuang52,oirsoy\}}@bloomberg.net
  }
}

\begin{document}
\maketitle

\begin{abstract}
    Retrieval over visually rich documents has a \emph{representation problem}: important content often lives in tables, charts, figures, and layout relations that plain OCR linearizes, corrupts, or omits.
    \emph{ColPali-family visual retrievers} address this with patch-level multi-vector indexes and late-interaction scoring, keeping image-derived retrieval on the query-time serving path.
    We introduce \midr{} (\textbf{M}ultimodal \textbf{I}ndexing for \textbf{D}ocument \textbf{R}etrieval), a training-free framework for \emph{enrichment-augmented indexing} that shifts multimodal reasoning to \emph{index time}. During ingestion, a multimodal LLM converts rendered pages into verified textual fields that are indexed with BM25F and optionally fused with dense retrieval, enabling text-centric serving over multimodally grounded evidence.
    On ViDoRe V3, \midr{} Hybrid achieves 0.6219 average \ndcg{} across five English domains, a 23.0\% relative gain over BM25, remaining competitive with ColQwen2.5.
    On two French-document domains, enrichment bridges English queries and French page text, lifting BM25 from 0.1532 to 0.5448 \ndcg{} and outperforming ColQwen2.5.
    Across all seven domains, \midr{} leads ColQwen2.5 on four while using $\sim$9$\times$ smaller index memory and approximately 2$\times$ lower query latency.
    These results establish index-time multimodal reasoning as a compelling accuracy--deployment alternative to serving-time visual late interaction.
    \end{abstract}

\section{Introduction}
\label{sec:intro}

Visually rich documents often express important information outside ordinary running text.
In enterprise reports, filings, manuals, slide decks, and scientific documents, key facts and claims may be encoded in tables, charts, figures, captions, visual grouping, or page layout~\citep{loison2026vidorev3,dong-etal-2025-mmdocir}: a table value depends on row and column headers, a chart trend on axes and legends, and a slide-level claim on visual grouping.
Plain OCR linearizes these structures into a flat character stream, discarding the cues that make their content searchable~\citep{zhang2025ocr}.

This matters because Retrieval-Augmented Generation (RAG) is increasingly used for question answering over large enterprise document collections~\citep{lewis2020rag,fan2024ragsurvey,ma2025visa}, and its effectiveness depends on whether the retriever surfaces the right evidence in the first place.
Even as context windows expand~\citep{liu2025longcontext}, retrieval remains necessary to reduce inference cost~\citep{izacard2021fid,karpukhin2020dpr} and to search collections that exceed practical context budgets~\citep{qiu2025incontext,liu2025sliding}.

The challenge is especially acute in specialized domains such as finance~\citep{chen-etal-2021-finqa}, legal analysis~\citep{kulkarni2026legal}, medicine~\citep{singhal2025medqa}, and education~\citep{alawwad2025textbook}, where critical evidence is frequently layout-dependent and visually structured~\citep{loison2026vidorev3,cho2025m3docvqa,dong-etal-2025-mmdocir}.
Effective retrieval over such documents therefore requires going beyond extracted text alone~\citep{cho2024m3docrag,tanaka2025vdocrag}.

The state-of-the-art response has been to retrieve over rendered page images directly.
ColPali~\citep{faysse2025colpali} adapts a PaliGemma backbone with a late-interaction retrieval head~\citep{khattab2020colbert}, encoding each rendered page into patch-level multi-vector representations scored with MaxSim.
ColQwen2.5 extends this ColPali-style recipe to a Qwen2.5-VL backbone and achieves strong open-weight performance on ViDoRe V3~\citep{loison2026vidorev3}.
This line of work establishes an important lesson: visually rich document retrieval requires multimodal understanding.
It also makes a particular design choice: because OCR text is insufficient, the retrieval stack should serve queries over rendered-page image representations using large visual multi-vector indexes and late-interaction scoring.

This design is powerful, but it keeps visual retrieval on the serving path.
That distinction matters for practical RAG deployments, where indexing and querying have different cost profiles.
Documents are typically indexed in offline ingestion pipelines, whereas the resulting index may be searched repeatedly by many users and increasingly by agentic workflows that issue multiple retrieval calls per request~\citep{dong2026doc}.
Multimodal reasoning performed during indexing can therefore be amortized over future queries, while serving retrieval over visual multi-vector indexes carries recurring query-time costs: large image-derived indexes, a compatible multimodal query encoder, and late-interaction scoring over candidates.

\begin{figure*}[!t]
    \centering
    \includegraphics[width=\textwidth]{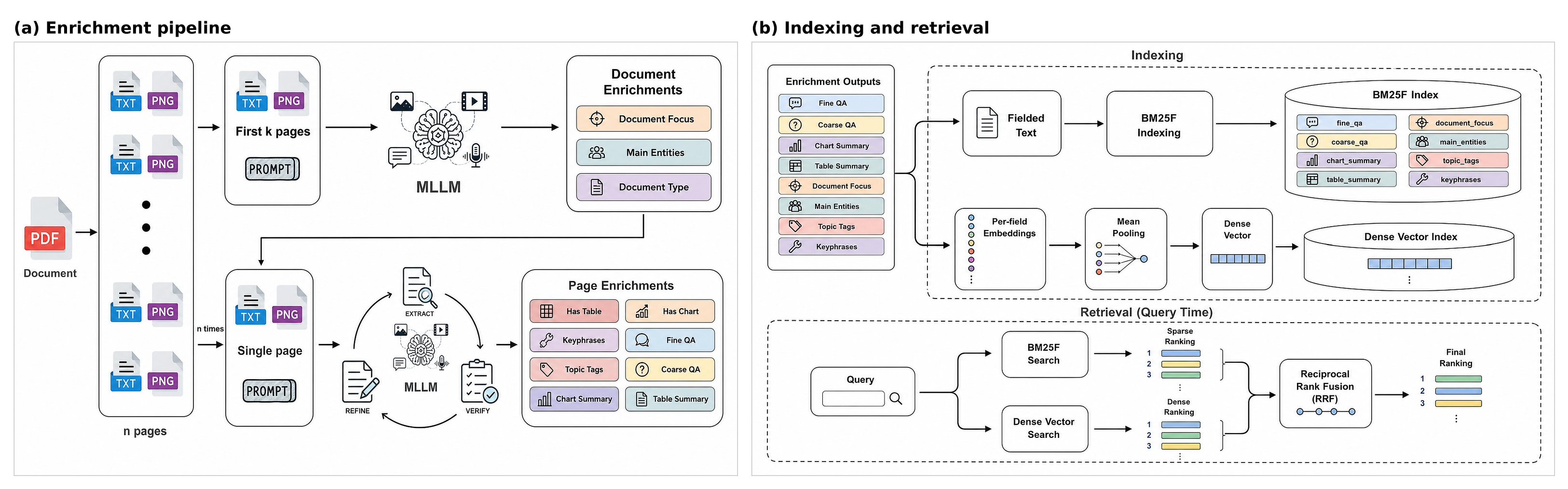}
    \caption{\textbf{\midr{} end-to-end pipeline.}
    \textbf{(a)} Enrichment pipeline: documents are decomposed into rendered page images and extracted page text; a document-level MLLM pass over the first five pages produces document enrichments, which condition page-level extract--verify--refine enrichment.
    \textbf{(b)} Indexing and retrieval: verified enrichment outputs are indexed as BM25F fields and dense vectors, then queried through lexical, dense, or hybrid retrieval with Reciprocal Rank Fusion, without serving-time visual multi-vector retrieval.}
    \label{fig:midr-full-pipeline}
\end{figure*}

Recent multimodal large language models (MLLMs) make it possible to revisit this design choice.
MLLMs can follow structured instructions, condition jointly on rendered page images and extracted text, and interpret layout-dependent evidence such as tables, charts, figures, and visually grouped content~\citep{zhang2024mm,team2026qwen3}.
This paper explores an alternative use of that capability: rather than serving every query over rendered-page image representations, use MLLMs during ingestion to convert visual and layout-dependent evidence into retrieval-ready textual fields.
\textit{The failure of OCR-only retrieval motivates multimodal document understanding; it does not necessarily require serving-time retrieval over rendered-page image representations.}

Much of the information a page provides to a retriever --- entities, claims, quantities, table structure, chart encodings, layout relations, captions, and visually grounded textual context --- is query-independent and a property of the document itself.
If these properties can be inferred during ingestion and materialized as structured textual evidence, then query-time retrieval can remain text-centric while still benefiting from multimodal document understanding.

We call this design \textbf{enrichment-augmented indexing}: use rendered pages for multimodal reasoning at index time, but serve retrieval through lexical, dense, and hybrid text-search infrastructure.

We introduce \textbf{\midr{}} (\textbf{M}ultimodal \textbf{I}ndexing for \textbf{D}ocument \textbf{R}etrieval), a training-free framework for enrichment-augmented indexing.
As shown in Figure~\ref{fig:midr-full-pipeline}, \midr{} separates enrichment from indexing and retrieval: document- and page-level enrichments are produced offline, then indexed as BM25F fields and dense vectors for text-centric serving.
A document-level pass first produces \emph{document enrichments}, including document type, primary focus, and main entities.
A page-level pass then uses these document enrichments together with each rendered page image and extracted page text to produce \emph{page-level enrichments}: layout and quality signals, table and chart descriptions, domain tags, keyphrases, and coarse- and fine-grained QA pairs.
Enrichment proceeds through an extract--verify--refine loop: page-level fields are first generated, then audited for grounding and consistency against the rendered page and extracted text, and finally revised only where verification identifies unsupported or inconsistent content.
The verified fields are then served through lexical, dense, and hybrid text retrieval (Section~\ref{sec:method}), so that at query time retrieval operates over multimodally grounded textual evidence, without processing page images or maintaining a visual multi-vector index.

On ViDoRe V3~\citep{loison2026vidorev3}, \midr{} Hybrid reaches 0.6219 average \ndcg{} across the five English-document domains, a 23.0\% relative gain over the raw-BM25 baseline of 0.5057, while remaining competitive with our ColQwen2.5 reproduction, with almost the entire remaining gap concentrated in computer science (Table~\ref{tab:main-results}).
On the two French-document domains, the same pipeline acts as a cross-lingual bridge: raw BM25 collapses to 0.1532 \ndcg{} because English queries share little surface vocabulary with French page text, while \midr{} Hybrid reaches 0.5448 and leads ColQwen2.5 on both domains by generating English enrichments at index time (Section~\ref{sec:cross-lingual}).
It does so from an index roughly 9$\times$ smaller than ColQwen2.5's visual multi-vector index, at 1.1--2.6$\times$ lower query latency depending on domain (Table~\ref{tab:efficiency-main}).
We additionally quantify the one-time ingestion cost and evaluate sensitivity to the enrichment MLLM.

The ablations indicate that \midr{} works by exposing distinct types of retrieval evidence rather than by adding undifferentiated generated text: QA pairs create lexical query--page bridges, keyphrases support dense semantic matching, and table summaries provide targeted gains on table pages, with the largest improvements where OCR loses the most structure (Tables~\ref{tab:group-ablation-main} and~\ref{tab:main-strata}).
A QA-only configuration nearly matches the full system on the English domains (0.6200 vs.\ 0.6219 \ndcg{}), while withholding the rendered page image reduces aggregate performance by 0.0113, with the effect concentrated in OCR-hard domains.
The comparison with visual retrieval also clarifies the boundary of the approach: multi-vector retrieval over rendered pages remains stronger on equations, diagrams, code layout, and fine-grained visual disambiguation, and a per-query oracle over the two systems reaches 0.7042 \ndcg{} (Appendix~\ref{app:oracle}), so the two encode complementary rather than redundant evidence.

\paragraph{Contributions and Key Findings}
\begin{itemize}[leftmargin=1.4em,itemsep=0.1em]
    \item We introduce \midr{} (\textbf{M}ultimodal \textbf{I}ndexing for \textbf{D}ocument \textbf{R}etrieval), a training-free framework for \emph{enrichment-augmented indexing}: a deployment alternative to ColPali-style visual retrieval that performs multimodal understanding over rendered pages at index time, then serves verified textual retrieval fields through BM25F, dense, and hybrid text-centric infrastructure.

    \item On ViDoRe V3, \midr{} Hybrid reaches 0.6219 average \ndcg{} on the English domains, a 23.0\% relative gain over raw BM25, remaining competitive with our ColQwen2.5 reproduction, from an index roughly 9$\times$ smaller than ColQwen2.5's.

    \item On the two French-document ViDoRe V3 domains, index-time enrichment acts as a cross-lingual bridge: \midr{} Hybrid lifts BM25 from 0.1532 to 0.5448 \ndcg{} by translating layout-grounded evidence into English at ingestion time and leads ColQwen2.5 on both (Section~\ref{sec:cross-lingual}).

    \item Field-, page-, query-, MLLM-, and oracle-level analyses, including QA-only and page-image ablations, show when enrichment helps and where visual retrieval remains stronger; a per-query oracle reaches 0.7042 \ndcg{}, indicating that index-time enrichment and visual late interaction encode complementary evidence types.
\end{itemize}
\section{Related Work}
\label{sec:related}

\midr{} builds on standard text retrieval infrastructure while addressing a representation gap that text-only retrievers cannot close by themselves.
BM25 and BM25F~\citep{robertson1994bm25,robertson2004bm25f} remain strong sparse baselines, dense bi-encoders~\citep{karpukhin2020dpr} improve semantic recall, and hybrid fusion such as Reciprocal Rank Fusion~\citep{cormack2009rrf} often combines their strengths.
Learned sparse models such as SPLADE~\citep{formal2021splade,formal2022spladev2} further expand textual matching surfaces.
These methods improve retrieval once evidence is present in the index, but they do not recover visual or layout-dependent evidence lost when OCR flattens tables, charts, and figures.

A related line of work enriches documents before indexing.
doc2query and docTTTTTquery~\citep{nogueira2019docexpansion} generate synthetic queries, Doc2Query++~\citep{kuo2025doc2query} improves coverage and fusion, and recent LLM-based systems such as EnrichIndex~\citep{chen2025enrichindex} and IndexRAG~\citep{baoshi2026indexrag} move query-independent reasoning offline through summaries, QA pairs, or bridging facts.
Closest to \midr{}, PREMIR~\citep{choi2025premir} uses a multimodal LLM to generate cross-modal pre-questions from documents before retrieval, and MLDocRAG~\citep{zhang2026mldocrag} generates fine-grained queries from heterogeneous multimodal chunks and links them across modalities and pages.
We therefore do not claim index-time enrichment as new.
\midr{} differs in that enrichment is \emph{fielded} and \emph{verified}: each page yields a typed multi-field record whose fields play distinct retrieval roles (Table~\ref{tab:schema}) rather than a single generated surface, and every field is audited against the rendered page and extracted text before indexing, with refinement firing on 9.6\% of pages overall and 52\% on the hardest domain.
We do not reimplement these systems as page-level ViDoRe retrievers: their retrieval units, pipelines, and target tasks differ from ours, so a reimplementation would not be the controlled comparison.
We instead report controlled simplifications of \midr{} itself---a QA-only variant, which is a doc2query-style single-surface configuration, and an OCR-only variant that withholds the page image (Section~\ref{sec:ablation}).

The dominant alternative is visual multi-vector retrieval.
ColBERT~\citep{khattab2020colbert} introduced late interaction over token-level vectors, and ColBERTv2~\citep{santhanam2022colbertv2} improved the efficiency and quality of this paradigm.
ColPali~\citep{faysse2025colpali} adapts late interaction to rendered document pages, and ColQwen2.5 extends the ColPali-style approach with a stronger vision-language backbone, setting the open source state of the art on ViDoRe V3~\citep{loison2026vidorev3}.
These systems address OCR failure by retrieving over rendered page images directly.
Optimized late-interaction kernels~\citep{pony2026flashmaxsim,sharma2026tilemaxsim} reduce the query-time cost of MaxSim scoring and would narrow the latency difference between the two designs; they do not reduce the memory needed to store patch-level page representations, which is where the designs differ structurally.
\midr{} accepts the need for multimodal understanding but separates it from query-time visual retrieval: rendered images are used during ingestion, while serving uses text-centric indexes.

Multimodal RAG systems such as M3DocRAG~\citep{cho2024m3docrag}, VDocRAG~\citep{tanaka2025vdocrag}, MDocAgent~\citep{han2025mdocagent}, and ViDoRAG~\citep{wang2025vidorag} combine document images, retrieval, agents, and generation, typically spending multimodal computation at query time.
Benchmarks such as REAL-MM-RAG~\citep{wasserman2025realmmrag}, MMDocIR~\citep{dong-etal-2025-mmdocir}, and ViDoRe V3~\citep{loison2026vidorev3} establish the need to evaluate retrieval over complex layouts and visually grounded evidence.
Query-side methods such as HyDE~\citep{gao2023hyde}, Guided Query Refinement~\citep{uzan2026gqr}, and multimodal reranking~\citep{geigle2022retrievefast} are complementary: they change how queries are processed, whereas \midr{} changes what the index contains.

\section{Enrichment-Augmented Indexing with \midr{}}
\label{sec:method}

\subsection{Problem Setup and Design Objective}
Let $\mathcal{D}$ be a collection of visually rich documents, where each document $d$ consists of pages $p_{d,i}$.
Each page has a rendered image $I_{d,i}$ and extracted text $x_{d,i}$, and the task is to rank pages for a text query $q$.
The design question is where multimodal computation should occur.
Visual multi-vector retrievers encode rendered pages into image-derived representations that remain on the query-time serving path.
In contrast, \midr{} derives compact document context $c_d$ and page-level enrichment fields $e_{d,i}$ during ingestion, then serves retrieval over textual fields $\{x_{d,i}, c_d, e_{d,i}\}$ using lexical, dense, or hybrid retrieval.
The objective is to amortize multimodal reasoning before queries arrive while preserving text-centric serving.

\midr{} implements this design as a training-free enrichment framework.
Given a visually rich document, it constructs a fielded textual representation of each page grounded in both the rendered page image and extracted page text.
As shown in Figure~\ref{fig:midr-full-pipeline}, \midr{} has three stages: document-level enrichment, verified page-level enrichment, and text-centric indexing.

\subsection{Document and Page Enrichments}
For each document, \midr{} first produces document-level enrichments from the first five pages.
These fields provide global context for page-level enrichment, helping disambiguate repeated entities, acronyms, and domain-specific references.
Each page is then enriched using four inputs: the rendered page image, extracted page text, document-level enrichments, and page metadata.
The image exposes layout, tables, charts, figures, and visual grouping that OCR may flatten; the extracted text preserves exact lexical evidence.
Table~\ref{tab:schema} lists the resulting document- and page-level enrichments, separating indexed retrieval fields from routing and control fields.

\begin{table}[t]
\centering
\scriptsize
\setlength{\tabcolsep}{2pt}
\renewcommand{\arraystretch}{0.86}
\begin{tabular}{p{0.12\columnwidth}p{0.28\columnwidth}p{0.50\columnwidth}}
\toprule
\textbf{Level} & \textbf{Enrichment} & \textbf{Role in \midr{}} \\
\midrule
\multicolumn{3}{l}{\textit{Indexed retrieval fields}} \\
\midrule
Document & \texttt{document\_focus}
& Global topic or purpose. \\

Document & \texttt{main\_entities}
& Salient organizations, products, datasets, regulations, drugs, or systems. \\

Page & \texttt{topic\_tags}
& Compact domain descriptors for lexical and dense matching. \\

Page & \texttt{keyphrases}
& Entity--metric--concept phrases for semantic matching. \\

Page & \texttt{table\_summary}
& Textualizes headers, units, rows, values, and comparisons. \\

Page & \texttt{chart\_summary}
& Textualizes axes, legends, trends, quantities, and visual relations. \\

Page & \texttt{coarse\_qa}
& Broad page-level QA pairs aligned with likely intents. \\

Page & \texttt{fine\_qa}
& Precise QA pairs for facts, values, definitions, cells, and visual details. \\
\midrule
\multicolumn{3}{l}{\textit{Routing and control fields}} \\
\midrule
Document & \texttt{document\_type}
& Genre or source type; conditions page interpretation. \\

Page & \texttt{layout}
& Page structure and visual content type; routes table/chart handling. \\

Page & \texttt{signal\_quality}
& Marks low-signal or decorative pages; gates enrichment. \\

Page & \texttt{verification\_issues}
& Unsupported or inconsistent fields found during verification. \\

Page & \texttt{refinement\_edits}
& Fields revised after verification for traceability. \\
\bottomrule
\end{tabular}
\caption{\midr{} enrichment schema. Indexed fields are used by BM25F and dense retrieval; routing and control fields guide enrichment, verification, and analysis.}
\label{tab:schema}
\end{table}


\subsection{Extract--Verify--Refine}
\label{sec:extract-verify-refine}

\begin{figure}[t]
    \centering
    \includegraphics[
        width=\columnwidth
    ]{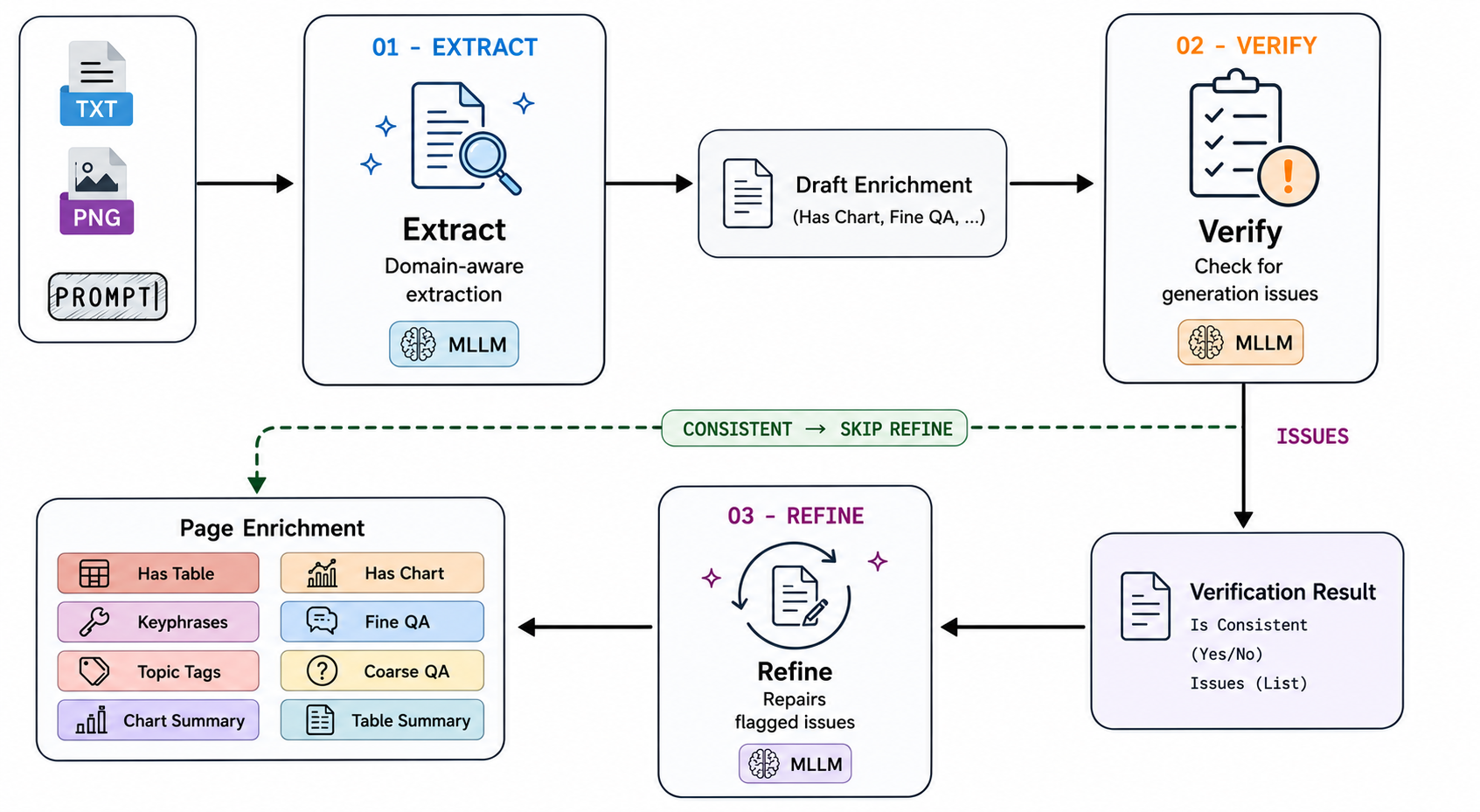}
    \caption{\textbf{Extract--verify--refine enrichment loop.}
    An initial structured enrichment is extracted from the page,
    verified against the rendered page image and extracted text,
    and selectively refined when verification identifies issues.}
    \label{fig:extract-verify-refine}
\end{figure}

Page enrichment follows an extract--verify--refine loop.
Figure~\ref{fig:extract-verify-refine} illustrates the three-stage
enrichment process.
The extractor generates a structured page enrichment from the rendered
page image, extracted text, document enrichments, and page metadata.
A deterministic postprocessor normalizes tags and keyphrases, removes
duplicate QA pairs, and enforces consistency between layout flags and
summary fields.
The verifier audits the draft against the rendered page and extracted
text for grounding, layout consistency, internal consistency, answer
quality, and completeness.
If issues are found, the refiner revises only the flagged fields and
postprocessing is applied again.
This conservative loop matters because index-time errors can affect
retrieval for many future queries.

\subsection{Indexing and retrieval}
The verified page enrichments and original page text are indexed as separate BM25F fields~\citep{robertson2004bm25f}.
Document-level indexed fields are replicated across the pages of the corresponding document so that each page exposes both local evidence and global context.
All BM25F fields are weighted uniformly; Appendix~\ref{app:implementation} reports an a-priori role-based variant and shows that the schema is insensitive to this choice.

For dense retrieval, \midr{} embeds the original page text and each enrichment field separately using EmbeddingGemma, then combines field embeddings with mean pooling.
For hybrid retrieval, \midr{} fuses BM25F and dense rankings with Reciprocal Rank Fusion~\citep{cormack2009rrf}.
Thus, multimodal reasoning is performed once during ingestion, while query-time retrieval operates over BM25F and dense text indexes containing multimodally grounded evidence, without page-image processing, visual multi-vector indexes, or late-interaction scoring.
\section{Experimental Setup}
\label{sec:setup}

\subsection{Benchmark}
We evaluate on \emph{ViDoRe V3}~\citep{loison2026vidorev3}, a page-level benchmark for multimodal document retrieval.
All experiments use the \emph{English-query configuration}, and all \midr{} enrichments are generated \emph{in English}, regardless of source-document language.
The primary evaluation covers five English-document domains---\emph{computer science}, \emph{finance}, \emph{HR}, \emph{industrial}, and \emph{pharmaceuticals}---with \textbf{1{,}489} queries over \textbf{12{,}968} pages from \textbf{101} documents.
We additionally evaluate two French-document domains---\emph{energy} and \emph{physics}---as a cross-lingual stress test, with \textbf{610} English queries over \textbf{3{,}899} pages from \textbf{83} documents.
Overall, the evaluation covers \textbf{2{,}099} queries over \textbf{16{,}867} pages from \textbf{184} documents; Appendix~\ref{app:coverage} gives the per-domain breakdown.
Following ViDoRe, the retrieval unit is a \emph{page}, the candidate pool is \emph{all pages within the corresponding domain}, and we report \ndcg{} using the official qrels and \texttt{ir-measures}~\citep{macavaney2022irmeasures}.

\subsection{Enrichment Model}
\midr{} uses the two-stage enrichment pipeline of Section~\ref{sec:method}, with GPT-5.1 as the enrichment MLLM for all main results.
We measure sensitivity to that choice by re-running the full pipeline with alternative backends and reindexing from scratch, holding retrieval, fusion, embeddings, and field weights fixed.
Frontier MLLMs cluster closely on the English evaluation, spanning 0.6120--0.6231 \ndcg{} with GPT-5.1 the strongest aggregate among them, while the open-weight backend we tested trails substantially and collapses on French (Appendix~\ref{app:mllm}).
GPT-5.1 is selected once on the five-domain English evaluation; we do not switch enrichment models by domain.

Page-level prompts are domain-aware across the five English-document domains and the two French-document domains.
The enrichment pipeline covers all \textbf{16{,}867} pages.
For the French-document domains, generating enrichments in English lets us test whether index-time enrichment can bridge English queries to French page content.

\subsection{Retrieval Systems}
We evaluate four text-centric retrieval configurations: \emph{BM25} over raw page markdown, \emph{BM25F} over enriched fields, \emph{dense retrieval} with mean-pooled EmbeddingGemma field embeddings, and \emph{\midr{} Hybrid}, which combines lexical and dense rankings.
We compare \midr{} against open-weight visual multi-vector retrievers: ColQwen2.5, the strongest ColPali-family model at comparable index size, and ColEmbed-3B-v2~\cite{moreira2026nemotron}, a stronger and substantially larger late-interaction retriever.
All retrieval-effectiveness numbers are produced by our own runs against the public qrels using the same corpora and evaluation code.
Dense and multi-vector experiments use FAISS~\citep{douze2025faiss} on a single NVIDIA L4.
Appendix~\ref{app:prior} documents the embedding-model and field-pooling choices behind the dense retriever and implementation details for the visual baselines.

\subsection{Ablations}
We evaluate group ablations, leave-one-out field ablations, and query-level stratifications.
Group ablations test compact variants such as \emph{QA-only}, \emph{no-QA}, \emph{semantic-only}, \emph{visual-only}, and \emph{no-semantic}.
Leave-one-out ablations measure the contribution of individual enrichment fields.
Stratified analyses group query results by visual content type and query type.
The full ablation suite and run inventory are provided in the appendix.

\section{Results and Analysis}
\label{sec:results}

We empirically evaluate the design claim behind \midr{}: \textbf{multimodal document understanding can be moved from query time to index time while preserving retrieval quality and reducing serving-time cost}.
We examine this claim step by step, moving from the main accuracy result and serving efficiency to cross-lingual behavior, enrichment mechanisms, and complementarity with visual multi-vector retrieval.

\noindent\textbf{How does index-time enrichment compare with visual multi-vector retrieval?}\par
\noindent Table~\ref{tab:main-results} gives the central result: \midr{} Hybrid reaches 0.6219 \ndcg{}, competitive with ColQwen2.5 aggregate of 0.6300 and leading on two of five English domains.
Relative to raw BM25 (0.5057), this is a 23.0\% gain, showing that much of the gap to visual retrieval can be closed through text-side representation alone, without serving-time visual matching.
The gain is not only a fusion effect: enriched BM25F reaches 0.5592, while dense mean-pool retrieval reaches 0.5898 over the same fielded representation, showing that multimodally grounded fields improve both lexical and semantic retrieval.

The aggregate gap is smaller than it first appears.
Computer science alone accounts for 0.045 of it (0.7170 vs.\ 0.7623), consistent with the advantage of visual encoders on formulaic notation, diagrams, and code layout.
Across the other four English domains the two systems average 0.5981 and 0.5969 respectively: \midr{} leads on HR (0.6043 vs.\ 0.6018) and pharmaceuticals (0.6424 vs.\ 0.6382), and trails by less than 0.002 on finance and industrial.

\begin{table*}[t]
\centering
\small
\setlength{\tabcolsep}{4.5pt}
\begin{tabular}{lcccccc}
\toprule
\textbf{System} & \textbf{CS} & \textbf{Finance} & \textbf{HR} & \textbf{Industrial} & \textbf{Pharma} & \textbf{Avg.} \\
\midrule
BM25 over markdown & 0.5774 & 0.4920 & 0.4851 & 0.4398 & 0.5341 & 0.5057 \\
Enriched BM25F     & 0.6497 & 0.5453 & 0.5292 & 0.4705 & 0.6012 & 0.5592 \\
Dense mean-pool    & 0.7107 & 0.6030 & 0.5474 & 0.4541 & 0.6341 & 0.5898 \\
ColQwen2.5         & \bestdom{0.7623} & \bestdom{0.6276} & 0.6018 & \bestdom{0.5200} & 0.6382 & \bestavg{0.6300}\\
\midrule
\textbf{\midr{} Hybrid (ours)}
& 0.7170 & 0.6261 & \bestdom{0.6043} & 0.5197 & \bestdom{0.6424} & 0.6219\\
\bottomrule
\end{tabular}
\caption{ViDoRe V3 English-domain \ndcg{}. Blue-shaded cells mark the best result within each domain; the orange-shaded cell marks the best aggregate. \midr{} is competitive with ColQwen2.5. \midr{} uses uniform field weights throughout.}
\label{tab:main-results}
\end{table*}

To broaden the comparison beyond ColQwen2.5, we additionally evaluate ColEmbed-3B-v2~\citep{moreira2026nemotron}, a stronger visual late-interaction retriever.
As shown in Table~\ref{tab:accuracy-index-tradeoff}, ColEmbed-3B-v2 reaches 0.6730 average \ndcg{} on the five English domains, establishing a higher accuracy point among the visual retrievers we evaluate.
This gain comes with a substantially larger visual multi-vector index: 11.07 MB per page compared with 0.038 MB per page for \midr{}.
The comparison therefore positions \midr{} as a strong accuracy--deployment operating point rather than the maximum-accuracy retriever.

\begin{table}[t]
\centering
\small
\setlength{\tabcolsep}{5pt}
\begin{tabular}{lcc}
\toprule
\textbf{System} & \textbf{EN-5 \ndcg{}} & \textbf{MB/page} \\
\midrule
\midr{} Hybrid          & 0.6219 & \textbf{0.038} \\
ColQwen2.5    & 0.6300 & 0.37 \\
ColEmbed-3B-v2          & \textbf{0.6730} & 11.07 \\
\bottomrule
\end{tabular}
\caption{Accuracy--index-size comparison on the five English-document ViDoRe V3 domains.
ColEmbed-3B-v2 achieves the highest retrieval accuracy, while \midr{} provides a substantially smaller text-centric index.}
\label{tab:accuracy-index-tradeoff}
\end{table}

\noindent\textbf{Does index-time enrichment change the serving-time cost profile?}\par
\noindent Table~\ref{tab:efficiency-main} summarizes the deployment tradeoff.
Normalized to BM25, ColQwen2.5 requires 27.9$\times$ query latency and 65.0$\times$ index memory, while \midr{} Hybrid uses 14.0$\times$ latency and 7.5$\times$ memory.
Thus, relative to ColQwen2.5, \midr{} Hybrid is roughly 2$\times$ faster and 9$\times$ smaller at query time.
ColEmbed-3B-v2 establishes a higher-accuracy operating point, but with a substantially larger visual multi-vector index (11.07 MB/page versus 0.038 MB/page for \midr{}).
Enriched BM25F alone provides an even cheaper operating point, at 3.4$\times$ latency and 2.5$\times$ memory.
This is the intended amortization tradeoff: \midr{} shifts multimodal computation into offline ingestion, whereas visual multi-vector retrieval keeps image-derived indexes, multimodal query encoding, and late-interaction scoring on the recurring query-time path.

\begin{table}[t]
\centering
\small
\setlength{\tabcolsep}{5pt}
\begin{tabular}{lcc}
\toprule
\textbf{Retrieval path} & \textbf{Latency} & \textbf{Memory} \\
                         & \multicolumn{2}{c}{\scriptsize lower is better} \\
\midrule
BM25              & 1.0$\times$  & 1.0$\times$  \\
Enriched BM25F    & 3.4$\times$  & 2.5$\times$  \\
\midrule
\ourscell{\midr{} Hybrid}
                  & \ourscell{14.0$\times$} & \ourscell{7.5$\times$} \\
\basecell{ColQwen2.5}
                  & \basecell{27.9$\times$} & \basecell{65.0$\times$} \\
\bottomrule
\end{tabular}
\caption{Average query latency and index memory normalized to BM25, measured with our retrieval implementations.
Green-shaded rows mark \midr{}; the orange-shaded row marks the visual multi-vector baseline.
\midr{} Hybrid is roughly 2$\times$ faster and 9$\times$ smaller than ColQwen2.5 at query time.}
\label{tab:efficiency-main}
\end{table}

\label{sec:cross-lingual}
\noindent\textbf{Can enrichment bridge language mismatch before retrieval?}\par
\noindent The French-document domains test this second consequence of enrichment-augmented indexing.
Here the documents are French and the queries are English.
Raw BM25 over French markdown collapses to 0.1532 average \ndcg{} (Table~\ref{tab:french-main}), since English queries share little surface vocabulary with French page text.
Visual retrievers handle this mismatch implicitly by matching rendered page images at serving time.
\midr{} handles it differently: because enrichments are generated in English from French page content, cross-lingual retrieval becomes monolingual matching against English enrichment fields at query time.
Enriched BM25F lifts the French average to 0.4606, and \midr{} Hybrid reaches 0.5448, exceeding reproduced ColQwen2.5 both on aggregate (0.5448 vs.\ 0.5315) and on each domain---energy (0.6192 vs.\ 0.5967) and physics (0.4704 vs.\ 0.4663)---without serving-time visual matching.
The language normalization happens once per document at ingestion, not once per query.

\begin{table}[t]
\centering
\small
\resizebox{\columnwidth}{!}{%
\begin{tabular}{lccc}
\toprule
\textbf{System} & \textbf{Energy} & \textbf{Physics} & \textbf{Avg.} \\
\midrule
BM25 over markdown    & 0.1577 & 0.1488 & 0.1532 \\
Enriched BM25F        & 0.5132 & 0.4081 & 0.4606 \\
Dense mean-pool       & 0.6042 & 0.4341 & 0.5192 \\
ColQwen2.5  & 0.5967 & 0.4663 & 0.5315 \\
\midrule
\textbf{\midr{} Hybrid (ours)}
    & \bestdom{0.6192} & \bestdom{0.4704} & \bestavg{0.5448} \\
\bottomrule
\end{tabular}%
}
\caption{French ViDoRe V3 \ndcg{} with English queries and French documents.
Blue-shaded cells mark best domains; orange-shaded cells mark best averages.
\midr{} uses English index-time enrichments to exceed reproduced ColQwen2.5 on aggregate and on both domains.}
\label{tab:french-main}
\end{table}

This result exposes a deployment property of text-mediated enrichment: the language of the index can be chosen during ingestion.
A multilingual collection can therefore be searched through a target-language textual index, while visual retrievers achieve language bridging only implicitly through rendered-page representations.
The result is specific to French documents with English queries; broader language-pair coverage remains future work.

\label{sec:ablation}
\label{sec:targeting}
\label{sec:robustness}
\noindent\textbf{What has enrichment actually added to the index?}\par
\noindent The 23.0\% aggregate lift over raw BM25 comes from partially non-overlapping lexical and dense contributions: Enriched BM25F reaches 0.5592 and dense mean-pool field retrieval 0.5898, which their RRF hybrid combines into 0.6219 (Table~\ref{tab:main-results}).
Group ablations in Table~\ref{tab:group-ablation-main} show that QA pairs are the dominant low-cost enrichment surface: \texttt{qa\_only} reaches 0.6200, recovering 94\% of the enrichment gain over a markdown-only hybrid baseline.
Conversely, removing QA drops hybrid retrieval to 0.5788, close to the markdown-only setting.
This supports the interpretation that generated QA pairs create the main query--page bridge, while the full schema adds targeted gains through table summaries, document focus, and other fields.

\begin{table}[t]
\centering
\small
\setlength{\tabcolsep}{5pt}
\begin{tabular}{lccc}
\toprule
\textbf{Configuration} & \textbf{BM25F} & \textbf{Dense} & \textbf{Hybrid} \\
\midrule
Markdown only          & 0.5057 & 0.5177 & 0.5737 \\
Semantic only          & 0.5057 & 0.5472 & 0.5843 \\
No QA                  & --     & 0.5606 & 0.5788 \\
QA only                & \textbf{0.5602} & 0.5622 & \underline{0.6200} \\
\midrule
\textbf{Full \midr{} (role-based)} & 0.5592 & \textbf{0.5898} & \textbf{0.6231} \\
\bottomrule
\end{tabular}
\caption{Group ablations on the five English domains.
Bold marks the best result per retrieval path; underline marks the compact QA-only hybrid result, which recovers most of the gain over markdown-only retrieval.}
\label{tab:group-ablation-main}
\end{table}

Field ablations point to distinct retrieval roles.
QA pairs provide the strongest lexical bridge, while keyphrases and QA fields support dense matching; full leave-one-out results appear in Appendix~\ref{app:field-ablations}.
Table summaries have little aggregate effect, but their value is concentrated on table pages, where removing them costs 0.028 \ndcg{}.

\noindent\textbf{Where does enrichment help most?}\par
\noindent Table~\ref{tab:main-strata} shows that the gains are largest where OCR loses structure.
Mixed-visual pages improve by 38.6\% relative \ndcg{}, table pages by 31.8\%, and numerical queries by 41.7\%.
Text-only pages and boolean queries improve least (+14.3\% and +11.7\%), where raw lexical evidence already gives BM25 a stronger starting point.
This supports the central mechanism: \midr{} helps most when retrieval depends on structure that OCR flattens.

\begin{table}[!t]
\centering
\scriptsize
\setlength{\tabcolsep}{3.2pt}
\renewcommand{\arraystretch}{0.88}
\begin{tabular}{lrrr}
\toprule
\textbf{Stratum} & \textbf{BM25} & \textbf{\midr{} Hyb.} & \textbf{$\Delta$} \\
\midrule
\multicolumn{4}{l}{\textit{Page visual content}} \\
Mixed visual  & 0.3957 & 0.5484 & +38.6\% \\
Table         & 0.4680 & 0.6170 & +31.8\% \\
Infographic   & 0.4867 & 0.6346 & +30.4\% \\
Chart         & 0.4993 & 0.6046 & +21.1\% \\
Text only     & 0.5618 & 0.6424 & +14.3\% \\
\midrule
\multicolumn{4}{l}{\textit{Query type}} \\
Numerical    & 0.4525 & 0.6414 & +41.7\% \\
Open-ended   & 0.3716 & 0.5136 & +38.2\% \\
Compare      & 0.5015 & 0.6208 & +23.8\% \\
Extractive   & 0.5896 & 0.7135 & +21.0\% \\
Multi-hop    & 0.4996 & 0.5711 & +14.3\% \\
Boolean      & 0.6004 & 0.6707 & +11.7\% \\
\bottomrule
\end{tabular}
\vspace{-1mm}
\caption{\midr{} Hybrid vs.\ raw BM25 by page visual content and query type on the five English domains.
$\Delta$ is relative \ndcg{} gain; full breakdowns are in Appendix~\ref{app:strata}.}
\label{tab:main-strata}
\end{table}

\noindent\textbf{Does the result depend on field weighting?}\par
\noindent ViDoRe~V3 provides no development split, so there is no principled way to tune BM25F field weights without fitting to the evaluation queries.
We therefore report \emph{uniform} weights, with every field set to 1.0, as \midr{}'s configuration throughout this paper, and we do not tune them.
As a check that this costs us nothing, we also evaluated a role-based weighting assigned a priori from each field's intended retrieval role (Appendix~\ref{app:implementation}): it scores 0.6231 on English, 0.0012 above uniform, and 0.5339 on French, 0.0109 below.
The two configurations are within noise on English and uniform is better on French, which indicates that the multi-field schema rather than weight tuning drives the gains.
Field boosts remain available as a per-deployment knob, but no headline result in this paper depends on setting them.

\label{sec:complementarity}
\noindent\textbf{Are enriched text retrieval and visual multi-vector retrieval substitutes?}\par
\noindent The paired analysis suggests they are not.
Because per-query scores are not available in the released ColQwen2.5 aggregates from~\citet{loison2026vidorev3}, this analysis uses our local ColQwen2.5 reproduction.
Across the 1{,}489 English-domain queries, roughly half have a decisive winner at the $>0.1$ \ndcg{} threshold: \midr{} wins 351 queries and ColQwen2.5 wins 362.
\midr{} tends to win when exact financial, pharmaceutical, or regulatory quantities have been verbalized into QA pairs and keyphrases, especially on pages whose tables would be flattened by raw OCR.
ColQwen2.5 tends to win when page evidence depends on distinctions that textual enrichment can blur, such as code blocks, equations, diagrams, or multiple visually similar tables.
A per-query oracle that selects the better of the two systems reaches 0.7042 \ndcg{}, a 13.0\% relative gain over \midr{} and 11.8\% over reproduced ColQwen2.5 (Appendix~\ref{app:oracle}).
The strongest reading is therefore \textit{complementarity, not replacement}: enriched text indexes and visual multi-vector indexes capture different signals.
ColEmbed-3B-v2 further shows that stronger visual retrieval can raise the accuracy ceiling, but at substantially greater index cost.

\section{Conclusion and Future Work}
\label{sec:conclusion}

\midr{} shows that multimodal reasoning for visually rich document retrieval can be moved from query time to index time.
By converting rendered pages into verified textual retrieval fields, enrichment-augmented indexing reaches performance comparable to strong visual multi-vector retrievers on ViDoRe V3 while exceeding ColQwen2.5 on the French domains, from an index roughly 9$\times$ smaller and at a one-time ingestion cost paid before any query arrives.
The same design enables index-time transformations such as generating English enrichment fields from French document pages, and ablations show that the gains are mechanistic: different fields support different retrieval paradigms, with improvements concentrated where OCR loses structure.
More broadly, \midr{} reframes multimodal retrieval as a systems-design question about \emph{where} multimodal understanding should occur, rather than \emph{whether} it is needed.

\midr{} is a framework rather than a fixed configuration: the enrichment schema, the prompts, the MLLM backend, the embedding model, and the field boosts are all deployment choices, and an improvement in any of them carries over without changing the serving path.
The broader conclusion is not that visual encoders are unnecessary, but that enriched text indexes and visual multi-vector indexes are complementary design points, and that index-time enrichment is worth reaching for when deployment constraints favor amortized ingestion, text-centric infrastructure, or auditable evidence.
A natural next step is adaptive retrieval: agents can route queries across typed enrichment fields---QA pairs, table summaries, keyphrases, and document context---and fall back to visual retrieval when fine-grained visual matching is required.

\section{Limitations}


\subsection{Comparison Scope}
Our evaluation covers ViDoRe V3, with detailed analysis on five English-document domains and a targeted cross-lingual study on two French-document domains.
We compare against open-weight visual retrievers reproduced locally under a single controlled pipeline, but ViDoRe V3 lists further systems that we do not evaluate, including API-only models for which controlled local measurement is not possible~\citep{loison2026vidorev3}.
\midr{} is also not the most accurate retriever in this space: ColEmbed-3B-v2 reaches 0.6730 average \ndcg{} on the English domains against \midr{}'s 0.6219, while storing roughly 11\,MB per page against \midr{}'s 0.038\,MB.
Our claim therefore concerns the accuracy--deployment tradeoff rather than peak accuracy, and a sweep of the full leaderboard would characterize that tradeoff more completely than the subset of systems we were able to run.

\subsection{Serving Cost}
The latency comparison is made against unoptimized late-interaction scoring.
Optimized MaxSim kernels~\citep{pony2026flashmaxsim,sharma2026tilemaxsim} reduce the query-time cost of visual multi-vector retrieval and would narrow this gap.
They do not reduce the cost of storing patch-level page representations, so index memory rather than latency is where the two designs differ structurally.

\subsection{Ingestion Cost}
\midr{} moves computation rather than removing it.
Ingestion requires on average 2.1 MLLM calls and approximately 8k tokens per page. This is a one-time index-building cost that is amortized over future queries, but it scales with corpus size and must be incurred again when documents change or the enrichment schema is revised.

\subsection{Enrichment MLLM}
Retrieval quality depends on the model used at index time.
Frontier backends cluster within 0.011 \ndcg{} on English, but the open-weight backend we tested trails by 0.046 on English and collapses on French (0.298 vs.\ 0.534; Appendix~\ref{app:mllm}).
The cross-lingual result in particular should not be assumed to transfer to every backend, and deployments restricted to open-weight models should expect a gap.

\subsection{Field and Language Coverage}
Chart summaries are net-neutral to negative on aggregate and clearly harmful on some domains, and main-entity strings retain source-language forms that act as confounders on French; both need prompt-level redesign before they generalize across source languages.
The cross-lingual study also covers one direction of one language pair.
The extract--verify--refine loop reduces unsupported and inconsistent fields before indexing, but grounding and coverage remain open challenges.



\bibliography{custom}
\clearpage
\appendix
\twocolumn[
\begin{center}
{\Large \bfseries Appendix}
\end{center}
\vspace{0.5em}
]
\section{Dataset and Enrichment Coverage}
\label{app:coverage}

\begin{table}[!htbp]
\centering
\small
\begin{tabular}{lrr}
\toprule
\textbf{Domain} & \textbf{Pages} & \textbf{Documents} \\
\midrule
Computer Science & 1,360 & 2 \\
Finance (EN) & 2,942 & 6 \\
HR & 1,110 & 14 \\
Industrial & 5,244 & 27 \\
Pharmaceuticals & 2,312 & 52 \\
Energy & 2,225 & 41 \\
Physics & 1,674 & 42 \\
\midrule
\textbf{Total} & \textbf{16,867} & \textbf{184} \\
\bottomrule
\end{tabular}
\caption{ViDoRe V3 per-domain page and document counts processed by
the GPT-5.1 extract--verify--refine pipeline; five English and two
French domains.}
\label{tab:coverage}
\end{table}

\section{Enrichment Schema and Field Weights}
\label{app:schema}

\begin{table}[!t]
\centering
\scriptsize
\setlength{\tabcolsep}{2pt}
\renewcommand{\arraystretch}{0.88}
\begin{tabular}{p{0.12\columnwidth}p{0.28\columnwidth}p{0.50\columnwidth}}
\toprule
\textbf{Level} & \textbf{Enrichment} & \textbf{Role in \midr{}} \\
\midrule
\multicolumn{3}{l}{\textit{Indexed fields}} \\
\midrule
Document
& \texttt{document\_focus}
& Primary topic or purpose; provides global context. \\

Document
& \texttt{main\_entities}
& Salient organizations, products, datasets, regulations, drugs, or systems; supports disambiguation. \\

Page
& \texttt{topic\_tags}
& Compact domain-specific descriptors for lexical and dense matching. \\

Page
& \texttt{keyphrases}
& Concise entity--metric--concept phrases for semantic matching. \\

Page
& \texttt{table\_summary}
& Textualizes rows, columns, headers, units, values, and salient comparisons. \\

Page
& \texttt{chart\_summary}
& Textualizes axes, legends, plotted quantities, trends, and visual relationships. \\

Page
& \texttt{coarse\_qa}
& Broad page-level QA pairs aligned with likely user intents. \\

Page
& \texttt{fine\_qa}
& Precise QA pairs for facts, values, definitions, table cells, and visual details. \\

\midrule
\multicolumn{3}{l}{\textit{Routing and control fields (not indexed)}} \\
\midrule
Document
& \texttt{document\_type}
& Document genre or source type; conditions page-level interpretation. \\

Page
& \texttt{layout}
& Page structure and visual content type; routes table/chart handling. \\

Page
& \texttt{signal\_quality}
& Indicates whether the page contains sufficient retrievable evidence or is low-signal/decorative; gates enrichment. \\

Page
& \texttt{verification\_issues}
& Unsupported or inconsistent fields identified during verification; used for refinement and auditing. \\

Page
& \texttt{refinement\_edits}
& Records fields revised after verification for traceability. \\
\bottomrule
\end{tabular}
\caption{\textbf{\midr{} enrichment schema.}
Document-level enrichments provide global context for page interpretation; page-level enrichments materialize visual and layout-dependent evidence as structured fields.
Indexed fields are matched against queries through BM25F and dense retrieval; routing and control fields support enrichment, verification, and analysis.}
\label{tab:schema-full}
\vspace{-1.5mm}
\end{table}

\section{Retrieval-Side Design Choices}
\label{app:prior}

This section reports the embedding backbone and field-pooling decisions used in the main results. We treat both as fixed design choices selected by preliminary comparison rather than as a primary axis of investigation; \midr{}'s contribution is the enrichment schema and the extract--verify--refine indexing path, not a new dense retriever.

\paragraph{Embedding backbones.}
We compared three open-weight text embedding models as the dense backbone for \midr{}:
\texttt{BAAI/bge-large-en-v1.5}~\citep{xiao2023bge}, a 335M-parameter English BERT-based encoder;
\texttt{Qwen/Qwen3-Embedding-0.6B}~\citep{zhang2025qwen3embed}, a 0.6B-parameter multilingual encoder with instruction-aware embeddings; and
\texttt{google/embeddinggemma-300m}~\citep{vera2025embedgemma}, a 308M-parameter multilingual encoder based on Gemma 3.
In preliminary English-5 comparisons, EmbeddingGemma produced the strongest dense-only \ndcg{} while matching or exceeding the other backbones on the French domains, where its multilingual training was a deciding factor. Because backbone selection is not the focus of this paper and the three models cluster within a narrow band on English, we report only the EmbeddingGemma configuration in the main results and leave a systematic embedding-model sweep to follow-up work.

\paragraph{Field-pooling variants.}
\midr{} embeds the original page text and each enrichment field separately, then combines the per-field embeddings into a single dense ranking. We compared four pooling strategies for combining query similarities across fields $i$, with field weight $w_i$ and per-field document embedding $e_{d_i}$, and $k$ and $r$ are the RRF parameters:
\begin{itemize}[noitemsep,topsep=0pt,leftmargin=1em]
   \item \textbf{weighted sum}:~$\sum_i w_i \cos(e_q, e_{d_i})$;  
    \item \textbf{max}:~$\max_i\left(w_i \cos(e_q, e_{d_i})\right)$; 
    \item \textbf{weighted RRF}:~$\sum_i w_i/(k + r_i)$;
    \item \textbf{mean pool}:~$\cos\left(e_q,\, \sum_i w_i e_{d_i} / \sum_i w_i\right)$.
\end{itemize}
In preliminary comparisons, mean pooling dominated the other three variants across both English and French domains. We adopt mean pooling as the default \midr{} dense representation throughout the main results and report no further pooling ablations, since the gap to the next-best variant was both small and consistent. As with the embedding backbone, we view pooling as a fixed design choice rather than a primary contribution.

\section{Group Ablations}
\label{app:group}

Tables~\ref{tab:group-ablation} and~\ref{tab:group-ablation-french}
report group ablations on the English and French domains. The English-French
asymmetry is sharp: on French, \texttt{qa\_only} matches the full
schema and \texttt{no\_semantic} stays close to it, while on English
the multifield schema yields larger targeted gains. Across all
seven domains, \texttt{qa\_only} reaches 0.5982 vs.\ 0.5976 hybrid
\ndcg{} for the full schema---non-QA fields contribute positively
on English but add offsetting noise on French
(Appendix~\ref{app:french}).

\begin{table}[!htbp]
\centering
\small
\begin{tabular}{lccc}
\toprule
\textbf{Configuration} & \textbf{BM25F} & \textbf{Dense} & \textbf{Hybrid} \\
\midrule
Full \midr{} (role-based) & 0.5592 & 0.5898 & 0.6231 \\
Markdown only & 0.5057 & 0.5177 & 0.5737 \\
QA only & 0.5602 & 0.5622 & 0.6200 \\
No QA & -- & 0.5606 & 0.5788 \\
Semantic only & 0.5057 & 0.5472 & 0.5843 \\
Visual only & 0.4620 & 0.4115 & 0.4779 \\
No semantic & -- & 0.5818 & 0.6156 \\
\bottomrule
\end{tabular}
\caption{Group ablations (English, 5 domains). Full \midr{} enrichment
(0.6231 hybrid) gains $+23.2\%$ over markdown-only BM25 (0.5057);
decomposed, enriched BM25F alone is $+10.6\%$ and dense mean-pool
alone is $+16.6\%$ (cf.\ Table~\ref{tab:main-results}). QA-only
captures most aggregate hybrid gains, while full enrichment gives
the best overall score and better targeted coverage.}
\label{tab:group-ablation}
\end{table}

\begin{table}[!htbp]
\centering
\small
\begin{tabular}{lccc}
\toprule
\textbf{Configuration} & \textbf{BM25F} & \textbf{Dense} & \textbf{Hybrid} \\
\midrule
Full \midr{} (role-based) & 0.4606 & 0.5192 & 0.5339 \\
Markdown only & 0.1532 & 0.4538 & 0.3357 \\
QA only & 0.4755 & 0.5177 & 0.5438 \\
No QA & -- & 0.4917 & 0.3842 \\
Semantic only & 0.1532 & 0.4854 & 0.3654 \\
Visual only & 0.2339 & 0.3752 & 0.3368 \\
No semantic & -- & 0.5136 & 0.5244 \\
\bottomrule
\end{tabular}
\caption{Group ablations (French, 2 domains). \texttt{qa\_only}
slightly exceeds the full schema (0.5438 vs.\ 0.5339 hybrid), and
\texttt{no\_semantic} stays close to the full schema---both consistent
with the cross-lingual sign flips in Appendix~\ref{app:french},
where non-QA fields contribute positively on English but add
offsetting noise on French.}
\label{tab:group-ablation-french}
\end{table}

\section{Stratified Field-Level Ablations}
\label{app:field-ablations}

\begin{table}[!htbp]
\centering
\small
\begin{tabular}{lrrr}
\toprule
\textbf{Removed field} & \textbf{BM25F} & \textbf{Dense} & \textbf{Hybrid} \\
\midrule
\texttt{fine\_qa}       & \textbf{-0.028} & -0.009          & \textbf{-0.016} \\
\texttt{coarse\_qa}     & -0.024          & \textbf{-0.011} & -0.014          \\
\texttt{keyphrases}     &  0.000          & -0.010          & -0.005          \\
\texttt{document\_focus} & -0.008          & -0.004          & -0.005          \\
\texttt{table\_summary} & +0.001          & -0.003          & -0.004          \\
\texttt{main\_entities} &  0.000          & +0.002          & -0.003          \\
\texttt{topic\_tags}    &  0.000          & -0.001          & -0.001          \\
\texttt{chart\_summary} & +0.007          & -0.004          & +0.003          \\
\bottomrule
\end{tabular}
\caption{Leave-one-out field ablations on English-5. Negative
deltas mean removing the field hurts. \textbf{Bold}: largest drop
per backend.}
\label{tab:loo}
\end{table}

Aggregate leave-one-out deltas (Table~\ref{tab:loo}) hide strongly
targeted gains. Field value is concentrated, not distributed. Table
summaries illustrate this sharply: aggregate hybrid delta 0.004
\ndcg{}, but on table pages specifically, removing them costs
0.028---a \textbf{seven-fold concentration}.
Table~\ref{tab:stratified} shows similar specialization: coarse QA
dominates numerical queries, fine QA dominates extractive ones, and
document focus helps open-ended and multi-hop queries where topical
alignment matters more than surface overlap.

\begin{table}[!htbp]
\centering
\small
\resizebox{\linewidth}{!}{%
\begin{tabular}{lll}
\toprule
\textbf{Field} & \textbf{Helps most on} & \textbf{Hurts most on} \\
\midrule
\texttt{table\_summary}   & table pages (-0.028)     & chart pages (+0.013) \\
\texttt{chart\_summary}   & infographics (-0.011)    & boolean queries (+0.021) \\
\texttt{coarse\_qa}       & numerical (-0.040)       & -- \\
\texttt{fine\_qa}         & extractive (-0.030)      & -- \\
\texttt{document\_focus}  & open-ended (-0.016)      & -- \\
\texttt{keyphrases}       & extractive (-0.018)      & chart pages (+0.010) \\
\bottomrule
\end{tabular}
}
\caption{Selected stratified leave-one-out findings on English-5.
Deltas are hybrid \ndcg{} changes when the field is removed (positive
means removing the field helps).}
\label{tab:stratified}
\end{table}

\paragraph{Negative findings for chart and entity fields.}
Chart summaries are the only field that hurts hybrid retrieval on
aggregate (+0.003 \ndcg{} when removed); \texttt{main\_entities}
shows a weaker version of the same pattern. We hypothesize that
broad trend descriptions (``increased over the period'',
``declining trajectory'') over-match boolean queries without
anchoring the correct page; the effect amplifies on French
(Appendix~\ref{app:french}). We treat both as deployment knobs to
toggle for cost-sensitive or multilingual settings, and leave
stricter value extraction and language-aware canonicalization to
follow-up work.

\section{Gains by Visual Content and Query Type}
\label{app:strata}

Per-stratum BM25 $\to$ BM25F $\to$ hybrid progressions backing the
aggregate in Table~\ref{tab:main-strata}. English strata in
Tables~\ref{tab:visual-strata}--\ref{tab:query-strata}; French in
Tables~\ref{tab:visual-strata-french}--\ref{tab:query-strata-french}.
Counts are query-level; small French strata (e.g., numerical $n=17$,
image $n=13$) should be read with sample size in mind.

\begin{table}[!htbp]
\centering
\small
\resizebox{\linewidth}{!}{%
\begin{tabular}{lrrrrr}
\toprule
\textbf{Visual content} & \textbf{n} & \textbf{BM25} & \textbf{BM25F} & \textbf{Hybrid} & \textbf{Total gain} \\
\midrule
Mixed visual & 222 & 0.3957 & 0.4832 & 0.5484 & +38.6\% \\
Table & 370 & 0.4680 & 0.5519 & 0.6170 & +31.8\% \\
Image & 25 & 0.5065 & 0.5768 & 0.6866 & +35.6\% \\
Infographic & 109 & 0.4867 & 0.5526 & 0.6346 & +30.4\% \\
Chart & 95 & 0.4993 & 0.5651 & 0.6046 & +21.1\% \\
Text only & 666 & 0.5618 & 0.5820 & 0.6424 & +14.3\% \\
\bottomrule
\end{tabular}
}
\caption{Stratified gains by visual content type (English, 5 domains).
Enrichment helps most where raw BM25 is weakest.}
\label{tab:visual-strata}
\end{table}

\begin{table}[!htbp]
\centering
\small
\resizebox{\linewidth}{!}{%
\begin{tabular}{lrrrrr}
\toprule
\textbf{Query type} & \textbf{n} & \textbf{BM25} & \textbf{BM25F} & \textbf{Hybrid} & \textbf{Total gain} \\
\midrule
Numerical & 49 & 0.4525 & 0.5393 & 0.6414 & +41.7\% \\
Open-ended & 309 & 0.3716 & 0.4272 & 0.5136 & +38.2\% \\
Compare-contrast & 260 & 0.5015 & 0.5578 & 0.6208 & +23.8\% \\
Extractive & 348 & 0.5896 & 0.6539 & 0.7135 & +21.0\% \\
Enumerative & 241 & 0.4839 & 0.5320 & 0.5922 & +22.4\% \\
Multi-hop & 91 & 0.4996 & 0.5459 & 0.5711 & +14.3\% \\
Boolean & 191 & 0.6004 & 0.6259 & 0.6707 & +11.7\% \\
\bottomrule
\end{tabular}
}
\caption{Stratified gains by query type (English, 5 domains).
Numerical and open-ended queries benefit most from index-time
enrichment.}
\label{tab:query-strata}
\end{table}

\begin{table}[!htbp]
\centering
\small
\resizebox{\linewidth}{!}{%
\begin{tabular}{lrrrrr}
\toprule
\textbf{Visual content} & \textbf{n} & \textbf{BM25} & \textbf{BM25F} & \textbf{Hybrid} & \textbf{Total gain} \\
\midrule
Infographic  & 28  & 0.1450 & 0.5775 & 0.6271 & +332.4\% \\
Table        & 79  & 0.1626 & 0.5257 & 0.6116 & +276.2\% \\
Mixed visual & 172 & 0.1283 & 0.3885 & 0.4586 & +257.5\% \\
Image        & 13  & 0.1378 & 0.5041 & 0.4758 & +245.2\% \\
Text only    & 213 & 0.1592 & 0.4554 & 0.5400 & +239.2\% \\
Other visual & 72  & 0.1470 & 0.4183 & 0.4908 & +233.8\% \\
Chart        & 33  & 0.2498 & 0.6998 & 0.7522 & +201.1\% \\
\bottomrule
\end{tabular}
}
\caption{Stratified gains by visual content type (French, 2 domains).
All strata show $>$200\% relative gains because raw BM25 on French
documents is near-floor (Section~\ref{sec:cross-lingual}); absolute
hybrid \ndcg{} values (0.46--0.75) are comparable in magnitude to
English (Table~\ref{tab:visual-strata}). The largest relative lifts
are on infographics and tables, where layout-dependent evidence is
most salient.}
\label{tab:visual-strata-french}
\end{table}

\begin{table}[!htbp]
\centering
\small
\resizebox{\linewidth}{!}{%
\begin{tabular}{lrrrrr}
\toprule
\textbf{Query type} & \textbf{n} & \textbf{BM25} & \textbf{BM25F} & \textbf{Hybrid} & \textbf{Total gain} \\
\midrule
Numerical        & 17  & 0.1579 & 0.6392 & 0.7254 & +359.5\% \\
Open-ended       & 152 & 0.0971 & 0.3587 & 0.4415 & +354.9\% \\
Enumerative      & 84  & 0.1196 & 0.4171 & 0.4769 & +298.9\% \\
Multi-hop        & 40  & 0.1156 & 0.3121 & 0.3963 & +242.9\% \\
Compare-contrast & 117 & 0.1744 & 0.5023 & 0.5663 & +224.6\% \\
Extractive       & 135 & 0.2069 & 0.5462 & 0.6317 & +205.3\% \\
Boolean          & 65  & 0.2010 & 0.5521 & 0.6039 & +200.5\% \\
\bottomrule
\end{tabular}
}
\caption{Stratified gains by query type (French, 2 domains). The
English ordering of relative gains (numerical and open-ended at the
top, boolean at the bottom; Table~\ref{tab:query-strata}) carries
over to French, but all magnitudes are inflated by the BM25 floor on
French documents. Sample sizes are small for the rarest strata
(numerical $n=17$, multi-hop $n=40$).}
\label{tab:query-strata-french}
\end{table}

\section{French Field Behavior}
\label{app:french}

The English leave-one-out picture (Table~\ref{tab:loo}) does not
transfer unchanged to French. Two shifts matter.

First, QA fields carry essentially all of the cross-lingual bridge
signal: removing \texttt{fine\_qa} costs 0.0315 \ndcg{} on French
versus 0.0162 on English, and \texttt{coarse\_qa} costs 0.0279
versus 0.0141 (Table~\ref{tab:french-loo}). Consequently,
\texttt{qa\_only} matches the full schema across all seven domains
(0.5982 vs.\ 0.5976 hybrid), exceeding the 94\% English-only
recovery.

Second, \texttt{chart\_summary} and \texttt{main\_entities} flip
sign: removing \texttt{chart\_summary} \emph{helps} French hybrid
retrieval by 0.0104 \ndcg{}, and removing \texttt{main\_entities}
helps by 0.0075. Chart descriptions generated from French visual
content produce noisy English representations that over-match
unrelated queries, and main-entity strings retain French forms or
transliterations that act as confounders. For multilingual
deployments, both fields likely need prompt-level redesign before
they generalize across source languages. Per-domain LOO deltas
appear in Table~\ref{tab:french-loo}; the flat zeros for
\texttt{keyphrases}, \texttt{topic\_tags}, and \texttt{main\_entities}
under BM25F reflect the role-based weighting from
Section~\ref{sec:method}, where these fields contribute essentially
through dense retrieval.

\begin{table*}[!t]
\centering
\small
\setlength{\tabcolsep}{4pt}
\begin{tabular}{lrrrrrrrrr}
\toprule
& \multicolumn{3}{c}{\textbf{BM25F}} & \multicolumn{3}{c}{\textbf{Dense}} & \multicolumn{3}{c}{\textbf{Hybrid}} \\
\cmidrule(lr){2-4} \cmidrule(lr){5-7} \cmidrule(lr){8-10}
\textbf{Removed field} & Energy & Physics & Avg & Energy & Physics & Avg & Energy & Physics & Avg \\
\midrule
\texttt{fine\_qa}        & \textbf{-0.0698} & \textbf{-0.0701} & \textbf{-0.0699} & -0.0090 & -0.0064 & -0.0077 & \textbf{-0.0353} & \textbf{-0.0278} & \textbf{-0.0315} \\
\texttt{coarse\_qa}      & -0.0707          & -0.0697          & -0.0702          & -0.0140 & -0.0042 & -0.0091 & -0.0343          & -0.0215          & -0.0279          \\
\texttt{keyphrases}      &  0.0000          &  0.0000          &  0.0000          & \textbf{-0.0151} & -0.0032 & \textbf{-0.0092} & -0.0027 & +0.0002 & -0.0012 \\
\texttt{document\_focus} & +0.0002          & -0.0053          & -0.0026          & -0.0024 & +0.0023 &  0.0000 & -0.0006 & +0.0054 & +0.0024 \\
\texttt{table\_summary}  & -0.0119          & +0.0027          & -0.0046          & -0.0021 & +0.0017 & -0.0002 & -0.0072 & +0.0068 & -0.0002 \\
\texttt{main\_entities}  &  0.0000          &  0.0000          &  0.0000          & +0.0082 & +0.0180 & +0.0131 & +0.0072 & +0.0078 & +0.0075 \\
\texttt{topic\_tags}     &  0.0000          &  0.0000          &  0.0000          & -0.0095 & -0.0131 & -0.0113 & -0.0043 & -0.0029 & -0.0036 \\
\texttt{chart\_summary}  & -0.0072          & +0.0259          & +0.0093          & -0.0050 & +0.0017 & -0.0017 & +0.0090 & +0.0118 & +0.0104 \\
\bottomrule
\end{tabular}
\caption{French leave-one-out field ablations. Negative deltas mean
removing the field hurts performance; positive deltas mean removing
the field helps. \textbf{Bold} marks the most consequential removal
per backend. Sign convention matches Table~\ref{tab:loo}.}
\label{tab:french-loo}
\end{table*}

\section{Per-Domain Latency and Memory}
\label{app:efficiency}

Table~\ref{tab:efficiency} decomposes Table~\ref{tab:efficiency-main} by domain,
reporting query latency and index memory normalized to BM25 on each domain.
Both tables report ratios rather than absolute measurements. The aggregate
column is the mean of the per-domain ratios, so it does not equal the ratio of
the corresponding per-domain means.

\begin{table*}[!t]
\centering
\scriptsize
\begin{tabular}{llrrrrrrrr}
\toprule
\textbf{Metric} & \textbf{Method} & \textbf{CS} & \textbf{Energy} & \textbf{Fin.} & \textbf{HR} & \textbf{Ind.} & \textbf{Phar.} & \textbf{Phys.} & \textbf{Avg.} \\
\midrule
\multirow{6}{*}{Latency}
& BM25 & 1.0 & 1.0 & 1.0 & 1.0 & 1.0 & 1.0 & 1.0 & 1.0 \\
& Enr. BM25F & 1.8 & 3.5 & 2.4 & 2.2 & 6.2 & 1.9 & 5.6 & 3.4 \\
& Gemma & 2.6 & 3.3 & 1.7 & 2.5 & 0.9 & 0.9 & 7.9 & 2.8 \\
& ColQwen2.5 & 17.1 & 32.3 & 22.2 & 13.7 & 23.0 & 26.7 & 60.5 & 27.9 \\
& Enr. BM25F+Gemma & 7.5 & 11.4 & 7.6 & 7.7 & 4.9 & 8.2 & 21.4 & 9.8 \\
& \midr{} Hybrid    & 8.8 & 14.1 & 9.2 & 8.9 & 20.2 & 10.3 & 26.4 & 14.0 \\
\midrule
\multirow{6}{*}{Memory}
& BM25 & 1.0 & 1.0 & 1.0 & 1.0 & 1.0 & 1.0 & 1.0 & 1.0 \\
& Enr. BM25F & 2.1 & 2.5 & 2.1 & 2.4 & 2.3 & 1.8 & 4.0 & 2.5 \\
& Gemma & 0.4 & 0.4 & 0.5 & 0.3 & 0.6 & 2.4 & 0.9 & 0.8 \\
& ColQwen2.5 & 56.0 & 43.6 & 59.0 & 42.7 & 76.7 & 72.4 & 105.0 & 65.0 \\
& Enr. BM25F+Gemma & 3.0 & 2.8 & 2.9 & 2.8 & 3.6 & 3.3 & 4.9 & 3.3 \\
& \midr{} Hybrid    & 6.6 & 5.6 & 6.7 & 5.5 & 8.4 & 7.9 & 11.7 & 7.5 \\
\bottomrule
\end{tabular}
\caption{Per-domain query latency and index memory, normalized to
BM25 on each domain. Aggregate values in the rightmost column match
Table~\ref{tab:efficiency-main}. Physics is the most expensive
domain for every method; ColQwen2.5's memory cost scales with page
count and visual complexity, ranging from 42.7$\times$ (HR) to
105.0$\times$ (physics), while \midr{} Hybrid stays within
5.5$\times$--11.7$\times$ across the same domains.
``Enr.\ BM25F+Gemma'' uses single-vector Gemma; \midr{} Hybrid uses
mean-pooled (MP) Gemma over enrichment fields.}
\label{tab:efficiency}
\end{table*}

\section{Offline Ingestion Cost}
\label{app:cost}

Table~\ref{tab:ingestion-cost} summarizes the offline cost of the
extract--verify--refine pipeline over all 16{,}867 ViDoRe V3 pages,
using GPT-5.1 as the enrichment MLLM; Table~\ref{tab:cost-per-domain}
decomposes it by domain. Figures are aggregated from the per-page
enrichment traces. We report MLLM calls, tokens, wall-clock time, and
refinement rate rather than a monetary total, because the dollar cost
depends on the provider and the pricing agreement under which the
pipeline is run. The cost is one-time and paid at index build, whereas
visual multi-vector retrieval pays GPU cost on every query for the life
of the index.

\begin{table}[!htbp]
\centering
\small
\begin{tabular}{lr}
\toprule
\textbf{Ingestion metric} & \textbf{Per page} \\
\midrule
MLLM calls                  & 2.10 \\
Input tokens                & 6{,}257 \\
Output tokens               & 1{,}914 \\
Wall-clock (serial)         & 23.2\,s \\
Refinement rate             & 9.6\% \\
\bottomrule
\end{tabular}
\caption{Offline ingestion cost of the extract--verify--refine pipeline,
averaged over all 16{,}867 ViDoRe V3 pages with GPT-5.1 as the
enrichment MLLM. This is a one-time index-build expense, amortized over
all future queries.}
\label{tab:ingestion-cost}
\end{table}

\begin{table}[!htbp]
\centering
\small
\setlength{\tabcolsep}{3.2pt}
\resizebox{\linewidth}{!}{%
\begin{tabular}{lrrrrr}
\toprule
\textbf{Domain} & \textbf{Pages} & \textbf{Calls/pg} & \textbf{In tok/pg} & \textbf{Out tok/pg} & \textbf{Refine} \\
\midrule
Computer Science & 1{,}360 & 2.07 & 5{,}474 & 1{,}443 & 6.6\% \\
Finance (EN)     & 2{,}942 & 2.03 & 6{,}446 & 1{,}765 & 2.7\% \\
HR               & 1{,}110 & 2.52 & 10{,}951 & 7{,}198 & 52.3\% \\
Industrial       & 5{,}244 & 2.08 & 5{,}817 & 1{,}590 & 7.9\% \\
Pharmaceuticals  & 2{,}312 & 2.06 & 5{,}335 & 1{,}243 & 5.8\% \\
Energy (FR)      & 2{,}225 & 2.06 & 6{,}698 & 1{,}661 & 6.1\% \\
Physics (FR)     & 1{,}674 & 2.11 & 5{,}516 & 1{,}333 & 11.2\% \\
\midrule
\textbf{Total}   & \textbf{16{,}867} & \textbf{2.10} & \textbf{6{,}257} & \textbf{1{,}914} & \textbf{9.6\%} \\
\bottomrule
\end{tabular}%
}
\caption{Per-domain ingestion cost. ``Refine'' is the fraction of pages on
which verification reported an issue and the refiner was invoked. HR is a
strong outlier at 52.3\%, which also makes it the most expensive domain per
page; finance, with clean extracted text, refines on 2.7\% of pages.}
\label{tab:cost-per-domain}
\end{table}

Amortized over all pages, the three pipeline stages cost 2{,}344 input and
1{,}441 output tokens for extraction, 3{,}489 and 290 for verification, and
424 and 183 for the conditional refinement call. Serial wall-clock time is
17.7\,s, 3.9\,s, and 1.7\,s respectively; the stage figures are rounded
independently, so they sum to slightly more than the measured 23.2\,s
per-page total in Table~\ref{tab:ingestion-cost}. Verification is therefore cheap in
output tokens but expensive in input tokens, because it re-reads the draft
enrichment alongside the page; refinement is the cheapest stage and runs on
fewer than one page in ten.

\section{Contribution of the Page Image}
\label{app:ocr-only}

To isolate what the rendered page image contributes over extracted text alone,
we re-ran the entire enrichment pipeline with the page image withheld from the
extractor, verifier, and refiner, holding everything else fixed. The
document-level enricher never reads page images---it operates on extracted page
text only---so the resulting pipeline is end-to-end text-only, and the
difference between the two runs is attributable to the page image.

Because the paired runs were evaluated on the same machine, we also report the
full-enrichment control from that machine alongside the cached values used
elsewhere in the paper; the device difference is at most 0.0043 \ndcg{} on any
domain, which is why cached results are used for the rest of the paper.

\begin{table}[!htbp]
\centering
\small
\resizebox{\linewidth}{!}{%
\begin{tabular}{lrrrr}
\toprule
\textbf{Domain} & \textbf{OCR-only} & \textbf{Full} & \textbf{Image $\Delta$@10} & \textbf{$\Delta$@5} \\
\midrule
Pharmaceuticals  & 0.6137 & 0.6483 & \textbf{+0.0346} & +0.0314 \\
Computer Science & 0.7061 & 0.7191 & +0.0131 & +0.0132 \\
HR               & 0.5953 & 0.6009 & +0.0055 & +0.0033 \\
Finance (EN)     & 0.6314 & 0.6338 & +0.0024 & +0.0059 \\
Industrial       & 0.5199 & 0.5206 & +0.0008 & +0.0046 \\
\textit{EN-5 avg} & \textit{0.6133} & \textit{0.6245} & \textit{+0.0113} & --- \\
\midrule
Energy (FR)      & 0.5943 & 0.6085 & +0.0142 & +0.0201 \\
Physics (FR)     & 0.4590 & 0.4637 & +0.0047 & +0.0005 \\
\bottomrule
\end{tabular}%
}
\caption{Hybrid \ndcg{} with and without the page image at enrichment time,
both measured on the same machine. The image contributes most where OCR
mangles content---chemical and scientific notation on pharmaceuticals, code
and garbled tables on computer science---and least on clean extracted text,
including industrial, which has the highest proportion of table pages in the
corpus. French domains are reported separately because the cross-lingual
bridge confounds the comparison.}
\label{tab:ocr-only}
\end{table}

Withholding the image is also a poor cost saving. On finance it reduces input
tokens by 18\% but leaves output tokens unchanged, while the verifier's
refinement rate rises from 2.7\% to 5.2\%: without visual grounding the
verifier finds more to correct, and the extra refinement calls claw back part
of the saving.

\section{Enrichment MLLM Sensitivity}
\label{app:mllm}

Our main results use GPT-5.1 as the enrichment MLLM. To assess how
much of \midr{}'s gain depends on this specific choice, we re-ran
the full extract--verify--refine pipeline with three alternative
models---GPT-5.4, Claude Sonnet 4.5, and the open-source
Qwen3-Omni-30B-A3B~\citep{xu2025qwen3omni}---and reindexed each
corpus from scratch. All other components (retrieval, fusion,
embeddings, weights) are held fixed.

\begin{table}[t]
\centering
\small
\setlength{\tabcolsep}{3.0pt}
\resizebox{\linewidth}{!}{%
\begin{tabular}{lcccc}
\toprule
\textbf{Domain} & \textbf{GPT-5.1} & \textbf{GPT-5.4} & \textbf{Claude S4.5} & \textbf{Qwen3-O.} \\
\midrule
\multicolumn{5}{l}{\textit{English domains}} \\
Computer Sci.    & \textbf{0.7178} & 0.7065 & 0.7078 & 0.693 \\
Finance          & 0.6322 & 0.6224 & \textbf{0.6351} & 0.574 \\
HR               & \textbf{0.5990} & 0.5921 & 0.5946 & 0.526 \\
Industrial       & \textbf{0.5212} & 0.5087 & 0.5154 & 0.466 \\
Pharma  & 0.6454 & 0.6305 & \textbf{0.6584} & 0.627 \\
\textit{EN-5 avg}     & \textit{0.6231} & \textit{0.6120} & \textit{0.6223} & \textit{0.5772} \\
\midrule
\multicolumn{5}{l}{\textit{French domains}} \\
Energy           & 0.6084 & 0.5997 & \textbf{0.6143} & 0.294 \\
Physics          & 0.4594 & 0.4485 & \textbf{0.4660} & 0.302 \\
\textit{FR-2 avg}     & \textit{0.5339} & \textit{0.5241} & \textit{0.5402} & \textit{0.298} \\
\bottomrule
\end{tabular}
}
\caption{\midr{} Hybrid \ndcg{} when the enrichment MLLM is varied,
holding the rest of the pipeline fixed. \textbf{Bold} marks the best
MLLM per row. The three frontier models cluster within roughly 2.5\%
relative on English-5; Qwen3-Omni-30B-A3B trails by a wider margin
and performs substantially worse on French.}
\label{tab:mllm-sensitivity}
\end{table}

\paragraph{Frontier MLLMs cluster tightly.}
On English-5, the three frontier models span 0.6120--0.6231 \ndcg{}
(2.5\% relative best-to-worst), well inside the BM25 $\to$ \midr{}
gain of 23.2\%. \midr{}'s improvement therefore comes principally
from the enrichment-augmented indexing design rather than from any
specific MLLM. GPT-5.1 leads on computer science, HR, and industrial;
Claude Sonnet 4.5 leads on finance and pharmaceuticals (the largest
single-domain margin in the table, +0.013 \ndcg{} over GPT-5.1) and
on both French domains. The aggregate ordering should be read with
some caution: prompting, decoding temperature, and structured-output
configuration can each shift any single model by amounts comparable
to the deltas in Table~\ref{tab:mllm-sensitivity}, and GPT-5.4's
slight regression relative to GPT-5.1 is plausibly within that
variance band rather than a stable ranking.

\paragraph{Open-source MLLM: verifier calibration.}
Qwen3-Omni-30B-A3B trails the frontier cluster by 4--5 points on
English-5 (0.5772 vs.\ 0.6120--0.6231, $-7.9\%$ relative to GPT-5.1)
and performs substantially worse on French at 0.298. Two factors compounded. Qwen3-Omni
had lower structured-output reliability in our setup (a non-trivial
fraction of calls returned malformed JSON, requiring retries). More
interestingly, its verifier was systematically over-strict relative
to its own extractor: borderline-but-supported fields were flagged
as unsupported, and the refiner shrank or dropped them, producing
indexes \emph{sparser} than the unverified drafts. GPT-5.1 showed
the opposite pattern, with verification typically expanding coverage
where extraction was conservative. The verifier in
extract--verify--refine therefore becomes a limiting factor
on enrichment volume, with tightness determined by verifier
calibration rather than by the loop itself. We treat the
Qwen3-Omni numbers as suggestive, but the qualitative
direction is consistent across our runs.

\paragraph{Implication.}
Extract--verify--refine does not require the same model in both
roles. Pairing a frugal open-source extractor with a stronger
verifier (or vice versa) is a natural way to trade ingestion cost
against enrichment quality without retraining. A systematic study
of mixed extractor/verifier configurations is left to future work.

\section{Per-Query Complementarity}
\label{app:complementarity}

Table~\ref{tab:query-complement} partitions the 1{,}489
English-domain queries by per-query \ndcg{} gap between \midr{}
Hybrid and our local ColQwen2.5 reproduction. Roughly half of queries
(48\%) have a decisive winner at the $>0.1$ \ndcg{} threshold, split
nearly evenly between \midr{} (351) and ColQwen2.5 (362). Only 6\%
fall within a marginal $<0.1$ gap; the rest both succeed (35\%) or
both fail (11\%).

\begin{table}[t]
\centering
\small
\resizebox{\linewidth}{!}{%
\begin{tabular}{lcc}
\toprule
\textbf{Region} & \textbf{Count} & \textbf{Interpretation} \\
\midrule
\midr{} wins by $>0.1$ \ndcg{}     & 351 & Enriched facts/QA expose evidence. \\
ColQwen2.5 wins by $>0.1$ \ndcg{}  & 362 & Visual matching preserves structure. \\
Both succeed ($>0.5$ \ndcg{})      & 522 & Shared easy/evident queries. \\
Both fail ($<0.3$ \ndcg{})         & 160 & Multi-page or cross-document gaps. \\
Marginal gap ($<0.1$ \ndcg{})      &  94 & Practically tied. \\
\bottomrule
\end{tabular}
}
\caption{Per-query partition of \midr{} Hybrid vs.\ reproduced
ColQwen2.5 across the 1{,}489 English-domain queries.}
\label{tab:query-complement}
\end{table}

\section{Oracle Complementarity with ColQwen2.5}
\label{app:oracle}

Per-domain oracle gains on English (Table~\ref{tab:oracle}) and
French (Table~\ref{tab:oracle-french}). French
oracle gains ($+0.08$ \ndcg{} over either system) match the English
pattern, indicating complementarity extends to the cross-lingual
setting.

\begin{table*}[!t]
\centering
\small
\resizebox{\textwidth}{!}{%
\begin{tabular}{lrrrrr}
\toprule
\textbf{Domain} & \textbf{\midr} & \textbf{ColQwen2.5} & \textbf{Oracle} & \textbf{Gain over \midr} & \textbf{Gain over ColQwen2.5} \\
\midrule
Computer Science & 0.7178 & 0.7623 & 0.8119 & +0.0941 & +0.0497 \\
Finance          & 0.6322 & 0.6276 & 0.7221 & +0.0898 & +0.0944 \\
HR               & 0.5990 & 0.6018 & 0.6825 & +0.0835 & +0.0808 \\
Industrial       & 0.5212 & 0.5200 & 0.5948 & +0.0736 & +0.0748 \\
Pharmaceuticals  & 0.6454 & 0.6382 & 0.7095 & +0.0641 & +0.0713 \\
\midrule
\textbf{Average} & \textbf{0.6231} & \textbf{0.6300} & \textbf{0.7042} & \textbf{+0.0810} & \textbf{+0.0742} \\
\bottomrule
\end{tabular}
}
\caption{Oracle complementarity analysis on the five English ViDoRe V3
domains. A per-query oracle that selects the better of \midr{} and
ColQwen2.5 reaches 0.7042 average \ndcg{}, +0.0810 over \midr{} alone
and +0.0742 over ColQwen2.5, indicating that the two systems
succeed and fail on different queries. The \midr{} column uses
role-based field weights, which differ from the uniform configuration
reported in Table~\ref{tab:main-results} by 0.0012 on the English
aggregate. Averages are simple means across the five domains.}
\label{tab:oracle}
\end{table*}

\begin{table*}[!t]
\centering
\small
\resizebox{\textwidth}{!}{%
\begin{tabular}{lrrrrr}
\toprule
\textbf{Domain} & \textbf{\midr} & \textbf{ColQwen2.5} & \textbf{Oracle} & \textbf{Gain over \midr} & \textbf{Gain over ColQwen2.5} \\
\midrule
Energy   & 0.6084 & 0.5967 & 0.6891 & +0.0807 & +0.0924 \\
Physics  & 0.4594 & 0.4663 & 0.5407 & +0.0813 & +0.0744 \\
\midrule
\textbf{Average} & \textbf{0.5339} & \textbf{0.5315} & \textbf{0.6149} & \textbf{+0.0810} & \textbf{+0.0834} \\
\bottomrule
\end{tabular}
}
\caption{Oracle complementarity analysis on the two French ViDoRe V3
domains. \midr{} and ColQwen2.5 are within
$\pm 0.012$ \ndcg{} of each other on both domains, and the oracle
gains ($+0.0810$ over \midr{}, $+0.0834$ over ColQwen2.5) are
comparable to the English case in Table~\ref{tab:oracle}---confirming
that the complementarity is not an artifact of either source
language.}
\label{tab:oracle-french}
\end{table*}

\section{Implementation Details}
\label{app:implementation}

We document the package versions and configuration parameters used
across the retrieval, embedding, and evaluation stack to support
reproduction.

\paragraph{Lexical retrieval (BM25 and BM25F).}
BM25 over markdown and BM25F over enriched fields are both
implemented using \texttt{Whoosh}'s \footnote{\url{https://github.com/Sygil-Dev/whoosh-reloaded}} scoring framework. Tokenization
uses Whoosh's default analyzer (lowercasing, standard stopword
removal, no stemming applied to enrichment fields to preserve named
entities and domain-specific tokens). BM25 parameters are left at
their library defaults ($k_1 = 1.2$, $b = 0.75$), and the same $k_1$ and $b$
apply per field for BM25F.

\paragraph{Field weights.}
All results in the main paper use uniform BM25F field weights, with every field
set to 1.0. ViDoRe V3 ships no development split, so tuning field weights would
mean fitting the evaluation queries; we therefore do not tune them.
As a robustness check we also evaluated a role-based weighting assigned a priori
from each field's intended retrieval role---page text 1.0, keyphrases and main
entities 1.2, document focus 1.1, topic tags 0.9, table and chart summaries 0.8,
coarse and fine QA fields 0.7. Table~\ref{tab:weights} compares the two.
The role-based scheme is 0.0012 better on the English aggregate and 0.0109 worse
on French, and no per-domain difference exceeds 0.011, so the schema is
insensitive to this choice. Field boosts remain a per-deployment knob for
practitioners who do have a validation set.

\begin{table}[!htbp]
\centering
\small
\begin{tabular}{lrrr}
\toprule
\textbf{Domain} & \textbf{Uniform} & \textbf{Role-based} & \textbf{$\Delta$} \\
\midrule
Computer Science & 0.7170 & 0.7178 & $+0.0008$ \\
Finance (EN)     & 0.6261 & 0.6322 & $+0.0061$ \\
HR               & 0.6043 & 0.5990 & $-0.0053$ \\
Industrial       & 0.5197 & 0.5212 & $+0.0015$ \\
Pharmaceuticals  & 0.6424 & 0.6454 & $+0.0030$ \\
\textit{EN-5 avg} & \textit{0.6219} & \textit{0.6231} & \textit{$+0.0012$} \\
\midrule
Energy (FR)      & 0.6192 & 0.6084 & $-0.0108$ \\
Physics (FR)     & 0.4704 & 0.4594 & $-0.0110$ \\
\textit{FR-2 avg} & \textit{0.5448} & \textit{0.5339} & \textit{$-0.0109$} \\
\bottomrule
\end{tabular}
\caption{Hybrid \ndcg{} under uniform and a-priori role-based BM25F field
weights. $\Delta$ is role-based minus uniform; positive means role-based is
better. Uniform is reported throughout the paper.}
\label{tab:weights}
\end{table}

\paragraph{Dense retrieval.}
Page text and each enrichment field are embedded with
EmbeddingGemma using the \texttt{transformers} library for model
loading and inference. Field embeddings are combined by mean
pooling and L2-normalized. The dense index is a FAISS
\texttt{IndexFlatIP} (exact inner-product search, no quantization
or approximate-search structures), built per domain over all pages
in the domain's candidate pool.

\paragraph{Multi-vector retrieval (ColQwen2.5 reproduction).}
For the local ColQwen2.5 reproduction used in the per-query
complementarity analysis (Section~\ref{sec:complementarity}), we
load the released checkpoint via \texttt{transformers} and score
candidates with the published late-interaction scoring routine.
Patch-level multi-vector representations are stored uncompressed.

\paragraph{Fusion.}
Hybrid retrieval combines BM25F and dense rankings using
Reciprocal Rank Fusion~\citep{cormack2009rrf} with the standard
constant $k = 60$.

\paragraph{Evaluation.}
All retrieval metrics are computed with
\texttt{ir-measures}~\citep{macavaney2022irmeasures} against the
official ViDoRe V3 qrels. We report \ndcg{} (the primary metric
defined by ViDoRe V3), evaluated at cutoff 10 across all systems
and configurations.

\paragraph{Enrichment generation.}
Multimodal LLM calls (GPT-5.1, GPT-5.4, GPT-5.4-mini, Claude
Sonnet 4.5, Qwen3-Omni-30B-A3B) use each provider's structured-output
mode where available, with low decoding temperature to favor
deterministic extraction.

%
%

\section{Enrichment Prompts}
\label{app:prompts}

This appendix documents the prompts used by the extract--verify--refine pipeline that produces \midr{}'s page-level
enrichments (Section~\ref{sec:method}). The pipeline runs three
stages per page: an \emph{extraction} call that generates a draft
enrichment from page text and image; a \emph{verification} call that
audits the draft against the page; and a conditional
\emph{refinement} call that fixes only the issues flagged by the
verifier. Each stage uses a system prompt and a user-message
template; the templates below show the prompts that produced all
results reported in this paper.

\subsection{Extraction prompts}
\label{app:prompts:extract}

The extraction system prompt is composed of a domain-specific header
followed by a shared rule block (Figure~\ref{fig:extract-common}).
We use eight domain headers: one per ViDoRe~V3 domain (finance,
computer science, energy, HR, industrial, pharmaceuticals, physics)
plus a \texttt{general} fallback. Two representative headers are
shown in Figures~\ref{fig:extract-finance}
and~\ref{fig:extract-physics}; the remaining six follow the same
structure (one paragraph of domain-specific guidance covering
terminology, units, table conventions, and topic-tag priorities) and
are available in the accompanying release. The user-message
template (Figure~\ref{fig:extract-user}) is shared across all
domains and is sent alongside the page image as a multimodal input.

\begin{promptfigure}
\begin{fullpromptcard}{Extraction - Shared rule block (appended to every domain header)}
CRITICAL RULES --- follow these exactly:

1. Use ONLY information explicitly present on THIS PAGE.

2. Document-level context (type, focus, entities) is for grounding only --- never invent facts from it.

3. If uncertain about a field, omit it or use defaults (empty list, "none").

4. Empty lists are always better than fabricated content.

LAYOUT DETECTION:

- TABLE means rows AND columns of structured data (e.g., financial statements, specification tables).

- Bullet points, numbered lists, and flowing text are NOT tables.

- CHART means any data visualization: bar chart, line graph, pie chart, scatter plot, diagram.

- Decorative images, logos, and photographs are NOT charts.

SIGNAL QUALITY:

- "high": Page contains substantive factual content useful for retrieval.

- "low": Title pages, blank pages, table of contents, decorative pages.

SUMMARIES:

- If has\_table is false, table\_summary MUST be "none".

- If has\_chart is false, chart\_summary MUST be "none".

- Summaries should capture: what the table/chart shows, key data points, structure (rows/columns/axes).

TOPIC TAGS:

- 5--10 lowercase tags describing the page's key topics.

- Use domain terminology, not generic words.

KEYPHRASES:

- Retrieval-optimized phrases a user might search for.

- Preserve exact values, units, entity names, and technical terms.

QA PAIRS:

- Coarse QA: High-level questions about the page's major insights (3--5 pairs).

- Fine QA: Specific questions about individual facts, one per data point (5--15 pairs).

- Answers MUST include value AND context --- never bare values.\\
\ \ BAD: "42\%" $\to$ GOOD: "Net revenue grew 42\% year-over-year to \$12.4B in Q4 2024."

- Include comparative questions when data spans time periods or entities.

- For tables: extract both totals and their sub-component breakdowns.

- No duplicate questions across coarse and fine sets.
\end{fullpromptcard}
\captionof{figure}{Shared rule block appended to every domain-specific header
in the extraction system prompt. Defines layout categories,
signal-quality labels, summary constraints, and QA structure.}
\label{fig:extract-common}
\end{promptfigure}

\begin{promptfigure}
\begin{fullpromptcard}{Extraction - Domain header: finance}
You are an expert document analyst specializing in financial documents.

DOMAIN-SPECIFIC GUIDANCE (FINANCE):

- Treat financial statements, metrics, and performance data as high-value factual content.

- Preserve exact values with currency symbols and units (e.g., \$12.4B, 15.3\%, \texteuro 500M).

- Recognize financial terminology: EBITDA, EPS, ROE, ROA, Basel III, Tier 1 capital, net interest income, provision for credit losses, tangible book value.

- Distinguish between reported results, forward guidance, and risk factors.

- For tables: capture line items, time periods, business segments, and comparative structure.

- Fine QA: ONE question per data cell; preserve exact currency values and units.

- Topic tags should prioritize: financial metrics, business segments, geographic markets, regulatory topics.
\end{fullpromptcard}
\captionof{figure}{Domain header for finance, prepended to the shared rule
block in Figure~\ref{fig:extract-common}. Other English-domain
headers (computer science, HR, industrial, pharmaceuticals) follow
the same structure.}
\label{fig:extract-finance}
\end{promptfigure}

\begin{promptfigure}
\begin{fullpromptcard}{Extraction - Domain header: physics (French source domain)}
You are an expert document analyst specializing in physics documents.

DOMAIN-SPECIFIC GUIDANCE (PHYSICS):

- Treat physical laws, principles, phenomena, and equations as high-value signals.

- Preserve mathematical formulas exactly (e.g., E = mc\^{}2, F = ma, $\nabla \times$B = $\mu_0$J).

- Preserve units and physical constants exactly (e.g., c = 3$\times$10\^{}8 m/s, $\hbar$, k\_B).

- Distinguish between theoretical derivations, experimental results, and problem examples.

- For tables: capture physical quantities, units, values, experimental conditions, uncertainties.

- Fine QA: ONE question per data cell; preserve exact values with units and uncertainties.

- Topic tags should prioritize: physical phenomena, physical laws/principles, particles/systems, experimental methods.

LANGUAGE: Generate ALL outputs in English, even if the source document is in another language. Translate content as needed while preserving exact values, units, and technical terms.
\end{fullpromptcard}
\captionof{figure}{Headers for French-source domains (physics, energy) and the
\texttt{general} fallback include an explicit \textsc{Language}
clause forcing English output for cross-lingual retrieval.}
\label{fig:extract-physics}
\end{promptfigure}

\begin{promptfigure}
\begin{fullpromptcard}{Extraction - User-message template (shared across domains)}
DOCUMENT CONTEXT:

- Document type: \{document\_type\}

- Document focus: \{document\_focus\}

- Key entities: \{main\_entities\}

PAGE TEXT:

\{page\_text\}

The PAGE IMAGE is also provided. Analyze both the text and image to produce all enrichment fields.
\end{fullpromptcard}
\captionof{figure}{Extraction user-message template. The three
document placeholders are filled from a separate
document-level enrichment pass (Section~\ref{sec:method}); the page
image is attached as a multimodal input alongside this text.}
\label{fig:extract-user}
\end{promptfigure}

\subsection{Verification prompt}
\label{app:prompts:verify}

After extraction, every draft enrichment is checked against the page
by a second call configured with the prompt in
Figure~\ref{fig:verify-system}. The verifier walks a five-point
checklist (layout consistency, fact grounding, internal consistency,
answer quality, completeness) and returns a structured list of
issues with field names, issue types, descriptions, and suggested
fixes. It sets \texttt{is\_consistent=true} only when zero issues
are found, in which case refinement is skipped. The user-message
template is in Figure~\ref{fig:verify-user}.

\begin{promptfigure}
\begin{fullpromptcard}{Verification - System prompt}
You are a meticulous verification agent. Your job is to check a draft page enrichment against the original page content (text + image) and identify any issues.

VERIFICATION CHECKLIST:

1. LAYOUT CONSISTENCY

\ \ - If has\_table=true, does the page actually contain a table (rows + columns)?

\ \ - If has\_chart=true, does the page actually contain a chart/visualization?

\ \ - If has\_table=false, is table\_summary set to "none"?

\ \ - If has\_chart=false, is chart\_summary set to "none"?

\ \ - Bullet points and lists are NOT tables.

2. FACT GROUNDING

\ \ - Is every claim in the enrichment supported by the page text or image?

\ \ - Are numerical values, dates, and entity names accurate?

\ \ - Are QA answers grounded in page content (not hallucinated)?

3. INTERNAL CONSISTENCY

\ \ - Do QA pairs contradict each other?

\ \ - Are there exact duplicate questions (in coarse\_qa, fine\_qa, or across both)?

\ \ - Do topic tags and keyphrases match the page's actual content?

4. ANSWER QUALITY

\ \ - Are answers contextual (value + context) or bare values?

\ \ - Bare values like "42\%" with no context are an issue.

5. COMPLETENESS

\ \ - Are obvious facts on the page missing from fine\_qa?

\ \ - For tables: is each cell accounted for?

For each issue found, specify:

- field: Which enrichment field has the issue

- issue\_type: One of "hallucination", "inconsistency", "missing\_context", "duplicate"

- description: What is wrong

- suggested\_fix: How to fix it

Set is\_consistent=true ONLY if zero issues are found.
\end{fullpromptcard}
\captionof{figure}{Verification system prompt. The five-point checklist
mirrors the constraints of the extraction prompt, letting the same
model audit drafts against the rules it was meant to follow.}
\label{fig:verify-system}
\end{promptfigure}

\begin{promptfigure}
\begin{fullpromptcard}{Verification - User-message template}
DRAFT ENRICHMENT (JSON):

\{draft\_json\}

PAGE TEXT:

\{page\_text\}

The PAGE IMAGE is also provided. Verify the draft enrichment against the page content.
\end{fullpromptcard}
\captionof{figure}{Verification user-message template. The draft enrichment
produced by extraction is passed back to the model verbatim,
alongside the original page text and image.}
\label{fig:verify-user}
\end{promptfigure}

\subsection{Refinement prompt}
\label{app:prompts:refine}

When verification flags one or more issues, a third call
applies targeted fixes using the prompt in
Figure~\ref{fig:refine-system}. Refinement is conditional: on pages
where verification returns no issues, this stage is skipped and the
draft enrichment is indexed as-is. The refiner is instructed to fix
only the listed issues and to log every change in a
\texttt{changes\_made} field, which feeds the
\texttt{refinement\_edits} routing field in
Table~\ref{tab:schema-full}. The user-message template is in
Figure~\ref{fig:refine-user}.

\begin{promptfigure}
\begin{fullpromptcard}{Refinement - System prompt (runs only when verification flags issues)}
You are a refinement agent. You receive a draft enrichment and a list of verification issues. Your job is to produce a corrected enrichment that fixes all identified issues while preserving everything that was correct.

RULES:

1. Fix ONLY the issues listed --- do not make unnecessary changes.

2. Preserve correct content exactly as-is.

3. When fixing QA answers, ensure they include value AND context.

4. When fixing layout issues, ensure summary fields are consistent with layout flags.

5. When removing duplicates, keep the more detailed/contextual version.

6. Log every change you make in the changes\_made list.

7. Generate all outputs in English, preserving exact values and technical terms.
\end{fullpromptcard}
\captionof{figure}{Refinement system prompt. The "fix only the listed issues"
constraint is what gives extract--verify--refine its surgical
behavior: the refiner is not free to rewrite the draft, only to
patch flagged fields.}
\label{fig:refine-system}
\end{promptfigure}

\begin{promptfigure}
\begin{fullpromptcard}{Refinement - User-message template}
DRAFT ENRICHMENT (JSON):

\{draft\_json\}

VERIFICATION ISSUES:

\{issues\_json\}

PAGE TEXT:

\{page\_text\}

The PAGE IMAGE is also provided. Fix the identified issues while preserving correct content.
\end{fullpromptcard}
\captionof{figure}{Refinement user-message template. The model receives both
the original draft and the structured list of issues from
verification, plus the page text and image for re-grounding.}
\label{fig:refine-user}
\end{promptfigure}

\section{Qualitative Analysis: How Enrichments Improve Retrieval}
\label{app:enrichment-examples}

To illustrate the mechanism by which \midr{} enrichments improve retrieval, we present two case studies drawn from the finance and computer science evaluation domains. In each case, we trace a query that the baseline BM25 system (operating on raw markdown text only) fails to retrieve correctly, and show how the enriched fields bridge the gap.

\subsection{Finance Domain: Vocabulary Mismatch on Restructuring Charges}
\label{app:ex-finance}

\paragraph{Query.}
\emph{``What were the total restructuring charges for the year 2020?''}

\paragraph{Gold page.}
Wells Fargo \& Company 2021 Annual Report (\texttt{NYSE\_WFC\_2021}), page~204.

\paragraph{Retrieval results.}
\begin{itemize}
  \item \textbf{Baseline BM25} (markdown only): target page \emph{not retrieved} in top~100 (nDCG@10\,=\,0.0).
  \item \textbf{Enriched BM25F} (with \midr{} fields): target page retrieved at rank~1 (nDCG@10\,=\,1.0).
\end{itemize}

\paragraph{Why the baseline fails.}
The page contains a detailed discussion of Wells Fargo's restructuring initiatives and an accrual table (``Accruals for Restructuring Charges''), but the raw markdown text is dominated by descriptions of personnel costs, facility closures, and accounting methodology. The specific phrase \emph{``total restructuring charges for the year 2020''} does not appear verbatim. Meanwhile, many other pages across the corpus mention ``restructuring charges'' in passing (e.g.,~Accenture, Texas Instruments, Nike), creating strong lexical competition from irrelevant documents.

\paragraph{How enrichments fix it.}
The \midr{} pipeline generates several enrichment fields that directly address this query:

\begin{itemize}
  \item \textbf{Coarse QA:} \emph{``What were Wells Fargo's total restructuring charges and their components for the year ended December 31, 2020?''} $\rightarrow$ \emph{``For the year ended December 31, 2020, Wells Fargo recorded total restructuring charges of \$726 million, including \$716 million of personnel costs\ldots''}

  \item \textbf{Table summary:} \emph{``Table~22.1, titled `Accruals for Restructuring Charges,' presents Wells Fargo \& Company's restructuring-related accrual activity by category (Personnel costs, Facility closure costs, Other, and Total) for the fiscal years ended December 31, 2019 and 2020, in millions.\ldots''}

  \item \textbf{Document focus:} \emph{``Wells Fargo \& Company 2021 financial performance and CEO discussion of strategic, risk, and operational transformation.''}
\end{itemize}

The coarse QA field is particularly effective here: it pre-generates a natural-language question that closely mirrors the user's query, creating the lexical overlap that the raw text lacks. The table summary further reinforces relevance by explicitly mentioning the year 2020 and the ``total'' category.

\paragraph{Extracted page text (abbreviated).}
{\small
\begin{verbatim}
The Company began pursuing various initiatives to reduce
expenses and create a more efficient and streamlined
organization in third quarter 2020. Actions from these
initiatives may include (i) reorganizing and simplifying
business processes [...] (ii) reducing headcount, (iii)
optimizing third-party spending [...]

Restructuring charges are recorded as a component of
noninterest expense on our consolidated statement
of income.
[...]

The following costs associated with these initiatives are
included in restructuring charges:
  - Personnel costs: Severance costs associated with
    headcount reductions [...]
  - Facility closure costs: Write-downs and acceleration
    of depreciation [...]
\end{verbatim}
}

\subsection{Computer Science Domain: Garbled Table Content}
\label{app:ex-cs}

\paragraph{Query.}
\emph{``How do range(1, 5) and range(1, 5, 2) differ in output pattern?''}

\paragraph{Gold pages.}
\emph{Introduction to Python Programming}, pages~135--136 and~149 (Chapter~5: Loops, \texttt{range()} function reference table). Three pages are marked relevant in the ground truth.

\paragraph{Retrieval results.}
\begin{itemize}
  \item \textbf{Baseline BM25} (markdown only): no target page in top~10; first target at rank~21 (nDCG@10\,=\,0.0).
  \item \textbf{Enriched BM25F} (with \midr{} fields): all three target pages at ranks~1, 2, and~3 (nDCG@10\,=\,1.0).
\end{itemize}

\paragraph{Why the baseline fails.}
The relevant pages contain a reference table (Table~5.1) that lists \texttt{range()} function call patterns with their outputs. However, the markdown extracted from the PDF via VLM-based conversion renders the table as garbled pipe-delimited text with OCR artifacts. The specific calls \texttt{range(1,\,5)} and \texttt{range(1,\,5,\,2)} do not appear in the markdown at all; the table shows different examples such as \texttt{range(4)}, \texttt{range(2,\,6)}, and \texttt{range(1,\,7,\,2)}. The word ``differ'' also does not appear. The baseline therefore cannot match this page to the query.

\paragraph{How enrichments fix it.}
\begin{itemize}
  \item \textbf{Table summary:} \emph{``The table titled `Using the range() function' lists three Python range() function call patterns (range(end), range(start, end), and range(start, end, step)), each with a textual description, one or more example calls, and the resulting integer sequences. It emphasizes that sequences start at 0 or the given start value, end before the end value, and use a specified step size (default 1 or custom).''} converts the garbled table into clean, searchable prose that introduces the key terms \emph{start}, \emph{end}, \emph{step}, and \emph{pattern}.

  \item \textbf{Fine QA:} \emph{``How does range(start, end, step) behave in terms of start, end, and step size according to the table?''} $\rightarrow$ \emph{``The form range(start, end, step) generates a sequence beginning at start until end with a step size equal to step.''} directly introduces the parametric framing that bridges to the user's query about how two-argument and three-argument calls ``differ.''

  \item \textbf{Fine QA:} \emph{``What output sequence is provided for the example call range(1, 7, 2)?''} $\rightarrow$ \emph{``The example range(1, 7, 2) produces the sequence 1, 3, 5.''} surfaces a concrete step-2 example with its output, providing lexical overlap with the query's \texttt{range(1, 5, 2)}.
\end{itemize}

The table summary and fine QA fields serve as a readable proxy for the tabular content that was lost during PDF-to-text conversion. The VLM that generates enrichments can interpret the table visually from the page image, recovering structured information that text extraction alone cannot.

\paragraph{Extracted page text (abbreviated).}
{\small
\begin{verbatim}
| Range() function in for loop | ...
|  | A for loop can be used for iteration and
    counting. The range() function is a common
    approach for implementing counting for loop.
    function of between the in a A range()
    generates a sequence integers two numbers
    given size. [...]
| Range function | Description | Example | Output |
| range (end) | Generates a sequence beginning
    at 0 until end. Step size: 1 | range (4) |
    0, 1, 2, 3 |
|  |  | range(0) 3) | 0, 1, 2 |
|  |  | range(2, 6) | 2, 3, 4, 5 |
| lange(start, end) | Generates a sequence
    beginning at start until end. Step size: 1 |
    range(-13, -9) | -13, -12, [...]
\end{verbatim}
}

\subsection{Discussion}

These examples illustrate two complementary mechanisms by which \midr{} enrichments improve lexical retrieval:

\begin{enumerate}
  \item \textbf{Vocabulary bridging.} The coarse and fine QA fields pre-generate natural-language questions that mirror how users formulate queries, bridging the gap between query vocabulary (e.g.,~``total restructuring charges,'' ``differ in output pattern'') and document vocabulary (e.g., accounting methodology prose, garbled table markup).

  \item \textbf{Structured content recovery.} The table summary field converts tabular and visual content, which is often poorly captured by PDF-to-text extraction, into clean, searchable prose. In the finance example, the table summary surfaces year-specific totals; in the computer science example, it recovers the semantics of a reference table that was garbled during text extraction.
\end{enumerate}

In both cases, the enrichments do not add new \emph{information}; the answers are present on the original pages. Instead, they re-express the page content in the vocabulary and structure that users naturally employ when searching. The finance example demonstrates vocabulary mismatch across documents, while the computer science example demonstrates information loss during text extraction from visually structured content.

\end{document}